\pdfoutput=1

\def\gsim{~\rlap{$>$}{\lower 1.0ex\hbox{$\sim$}}}
\def\lsim{~\rlap{$<$}{\lower 1.0ex\hbox{$\sim$}}}

\def\h2o{\rm{H_{2}O}}
\def\mh2{\rm{H_{2}}}

\def\co2{\rm{CO_{2}}}
\def\ch4{\rm{CH_{4}}}
\documentclass[12pt, preprint]{aastex}
\usepackage{natbib,color,lscape}
\usepackage{subfigure}
\usepackage{threeparttable}
\usepackage{url}
\usepackage{subfigure}
\usepackage{threeparttable}
\usepackage{rotating}

\begin{document}
\title{THE INNER EDGE OF THE HABITABLE ZONE FOR SYNCHRONOUSLY ROTATING PLANETS AROUND LOW-MASS STARS USING
GENERAL CIRCULATION MODELS }
\author{Ravi kumar Kopparapu\altaffilmark{1,2,3,4,5}, 
        Eric T. Wolf\altaffilmark{6}
        Jacob Haqq-Misra\altaffilmark{3,5},
        Jun Yang\altaffilmark{7}
        James F. Kasting\altaffilmark{1,3,4},
        Victoria Meadows\altaffilmark{3,9},
        Ryan Terrien\altaffilmark{4,8},
        Suvrath Mahadevan\altaffilmark{4,8}}

\altaffiltext{1}{Department of Geosciences, Penn State University, 443 
Deike Building, University Park, PA 16802, USA}
\altaffiltext{2}{New institution: NASA Goddard Space Flight Center, 8800 Greenbelt Road, Mail Stop 699.0
Building 34, Greenbelt, MD 20771}

\altaffiltext{3}{NASA Astrobiology Institute's Virtual Planetary Laboratory, P.O. Box 351580, Seattle, WA 98195, USA}
\altaffiltext{4}{Center for Exoplanets \& Habitable Worlds, The Pennsylvania State University, University
Park, PA 16802}
\altaffiltext{5}{Blue Marble Space Institute of Science, 1001 4th Ave, Suite 3201, Seattle, Washington 98154, USA}
\altaffiltext{6}{Department of Atmospheric and Oceanic Sciences, Laboratory for Atmospheric and Space Physics, University of Colorado
Boulder, Boulder, Colorado, USA}
\altaffiltext{7}{Department of Geophysical Sciences, University of Chicago}
\altaffiltext{8}{Department of Astronomy \& Astophysics, The Pennsylvania State University, 525 Davey
Laboratory, University Park, 16802, USA}
\altaffiltext{9}{Astronomy Department, University of Washington, Box 351580, Seattle, WA 98195-1580, USA}

\begin{abstract}
Terrestrial planets at the inner edge of the habitable zone of late-K and M-dwarf stars are expected to be in synchronous rotation, as a consequence of strong tidal interactions with their host stars. 
Previous global climate model (GCM) studies have shown that, for slowly-rotating planets, strong convection at the substellar 
point can create optically thick water clouds, increasing the planetary albedo, and thus stabilizing the 
climate against a thermal runaway. However these studies did not use self-consistent 
orbital/rotational periods for synchronously rotating planets placed at different 
distances from the host star. 
Here we provide new estimates of the inner edge of the habitable zone for synchronously rotating terrestrial 
planets around late-K and M-dwarf stars using a 3-D Earth-analog GCM with self-consistent relationships 
 between stellar metallicity, stellar effective temperature, and the planetary orbital/rotational period.
We find that both atmospheric dynamics and the efficacy of the substellar cloud deck are sensitive to the precise 
rotation rate of the planet. 
Around mid-to-late M-dwarf stars with low metallicity, planetary rotation rates at the inner edge of the HZ become faster,
and the inner edge of the habitable zone is farther away from the host stars than in previous GCM studies.
  For an Earth-sized planet, the dynamical regime of the substellar clouds begins to transition as the rotation 
rate approaches 
$\sim10$ days.  These faster rotation rates produce stronger zonal winds that encircle the planet and smear the 
substellar clouds around it, lowering the planetary albedo, and causing the onset of the water-vapor
greenhouse 
climatic instability  to occur at up to $\sim 25\%$ lower incident stellar fluxes than found in previous GCM 
studies. For mid-to-late M-dwarf stars with high metallicity and for mid-K to early-M stars, we agree with
previous studies.

\end{abstract}

\maketitle

\section{Introduction}
\label{intro}

         The study of habitable zones (HZs) has received increased attention 
with the discoveries of terrestrial mass/size planets from both ground-based surveys and from the Kepler
mission \citep{Kasting1993, Selsis2007b, Kopp2013, Leconte2013, Yang2013, Kopp2014, WT2014, Yang2014a,
WT2015}.
 Specifically, terrestrial planets in the HZs of M-dwarf stars have received the most recent
scrutiny because their shorter orbital periods increase their chances of detection compared to planets around G
stars. M-dwarfs are the most numerous stars in the Galaxy and are also our closest neighbors. 
The
upcoming James Webb Space Telescope ({\it JWST}) mission, with targets provided by the ongoing
K2 mission and planned Transiting Exoplanet Survey Satellite ({\it TESS}, \cite{Ricker2014}), {\it may} be 
capable of probing
the atmospheric composition of terrestrial planets around a nearby M-dwarf. 
This will be our first opportunity to obtain spectral information on terrestrial planets in the habitable zones of 
M-dwarf stars.
 The initial
targets for these missions will likely be planets near the inner edge of the HZ, precisely because such planets
have an increased probability of transit and are more likely to get their masses estimated by RV techniques.
 Recent work (Boyajian et al. 2012, Boyajian et al. 2014, Terrien et al. 2015, Mann et al. 2015,
Newton et al. 2015) has resulted in better stellar characterization and more accurate
temperatures and radii for the M dwarfs. Thus, determining an accurate inner edge of the HZ
around late-K and M-dwarf stars will soon become crucial for interpreting the data from these missions.

Previous 1-D climate model studies of HZs \citep{Kasting1993, Selsis2007b, Kopp2013, Kopp2014} predicted
that, for a water-rich planet such as Earth, two types of habitability limits exist near the inner edge:
(1) A moist greenhouse limit, which occurs when the stratospheric water vapor volume mixing
ratio becomes $> 10^{-3}$, causing the planet to lose water by photolysis and subsequent loss of
hydrogen to space; (2) A runaway greenhouse limit, whereby the outgoing thermal-infrared radiation from
the planet reaches an upper limit beyond which the surface temperature increases uncontrollably,
causing the oceans to evaporate. Habitable climates may be terminated via the moist greenhouse
process long before a thermal runaway occurs. Widespread surface habitability for human life 
could be terminated by rising temperatures well before even a moist greenhouse is reached, because of limits on biological functioning \citep{SH2010}.

1-D climate models that are being used currently have significant limitations. In particular, it is
hard to accurately model planets within the HZs of late-K and all
M-dwarf stars because such planets are expected to be tidally locked. If the planet's orbital eccentricity is near zero, 
this can result in synchronous rotation, in which one side of
the planet always faces the star. Indeed, a recent study by \cite{Leconte2015} argued that all planets near the inner edge 
of the HZ of M-dwarfs (masses $<0.6$ M$_{\odot}$; \cite{Dotter2008}) should rotate synchronously. 
 Somewhat surprisingly, according to 
 \cite{Leconte2015}, planets orbiting in the outer HZs of M and K-dwarf stars should not rotate synchronously because their rotation rates are perturbed by thermal tides.

Recently, \cite{Yang2013, Yang2014a} used the
Community Atmosphere Model (CAM) from the National Center of Atmospheric Research, Boulder CO,
 to simulate an Earth-size water-rich planet
in various spin-orbit configurations. 
Not surprisingly, they found that atmospheric circulation, and thus the predominant location of clouds, is critically influenced by the Coriolis force 
(which results from the planetary rotation rate).  Rapidly rotating planets, like Earth, have latitudinally banded circulation patterns.
However, for slowly rotating planets the Coriolis force is too weak to form such banded structures. Instead, strong 
and persistent convection occurs at the substellar region, creating a stationary and optically thick cloud deck.
This causes a strong increase in the planetary albedo, cooling the planet, and stabilizing climate against 
a thermal runaway for large incident stellar fluxes. 
This result was found to be qualitatively robust for a wide variety of model configurations and parameter 
sensitivity tests.  It has also been qualitatively corroborated by a generalized version of the NASA GISS GCM 
\citep{Way2015}.  Thus, it initially appears that the substellar convection and cloud mechanisms on slow rotators is a robust prediction of climate models.
This moves the inner edge of the habitable zone significantly 
closer to the star, reaching almost twice the stellar flux predicted by 1-D models \citep{Kopp2013, Kopp2014}.

A caveat should be added to the above conclusion, however.  In doing their calculations, \cite{Yang2014a} assumed a constant orbital period of 60 days at the inner edge of the HZ for synchronously 
rotating planets around M and K-dwarf stars. While this assumption is sufficient to demonstrate the concept of cloud stabilization, 
it is not consistent with Kepler's third law. 
Moreover, the climatic differentiation between ``rapid'' and ``slow'' rotators is not a binary choice, but rather is a continuous 
function of changing Coriolis force \citep{Yang2014a}. Thus, correcting the orbital/rotational period for synchronously 
rotating planets around M and K-dwarfs may indeed change our view of the inner edge of the HZ.

Whether atmospheric circulation of one planet is in the rapidly rotating or slowly rotating regime mainly depends on the ratio of 
Rossby deformation radius to the planetary radius (Edson et al. 2011; Showman et al. 2013; Carone2014).
 If the ratio is larger than one, the 
planet will be in the slow rotating regime, and if the ratio is smaller than one, the planet will be in the rapidly rotating regime. 
For the synchronously rotating planets we disussed here, the transition from the slowly rotating regime to the rapidly rotating regime occurs when the 
planetary rotation period is approximately 5 days for planets with $1$ R$_\oplus$ and approximately 10 days for planets with 
2 R$_\oplus$, as pointed out by Yang et al. (2013). Different planets in one of the two regimes will have very similar patterns 
of atmospheric circulation and surface climate.

 For synchronously rotating planets, the orbital period and the rotational period are by definition in 1:1 resonance.
For example, 
Gl 667Cc \citep{Anglada-Escude2013}, a super-Earth around an M-dwarf star with effective temperature $T_{eff} = 3350$K,
has an orbital period of $28.1$ days \citep{Robertson2014}\footnote{exoplanets.org}. This planet receives a stellar flux of 
$0.87$ times Earth's flux. The model planets at the inner edge of the HZ with a $60$ day orbital period 
considered by \cite{Yang2014a} would need to be at roughly
twice the observed period of Gl 667Cc to be consistent with Kepler's third law.
 Therefore, they should be receiving roughly $\sim 0.3$ Earth's flux,
 which puts those model planets beyond the outer edge of the HZ, rather than at the inner HZ limit.
Planets at the inner edge of the HZ of M-dwarfs have shorter orbital periods than planets around G-dwarfs.
Since such planets are synchronously rotating, shorter orbital periods  mean shorter rotational periods.
As \cite{Yang2014a} pointed out, a rapidly rotating planet tends to have decreased substellar cloud cover, lowering  
its albedo, and hence moving the inner edge of the HZ further away from the star.
Our goal in this paper is to use self-consistent orbital and rotational periods for planets and to derive new inner HZ limits for synchronously 
rotating planets around  M and late-K dwarfs. 

The outline of the paper is as follows: In \S\ref{sec2} we briefly describe
	our general circulation model.
	In \S\ref{results} we present results from our  model, comparing to previous studies, and present new HZ limits.
We conclude in
	\S\ref{conclusions}.

	\section{Model}
         We used the Community Atmosphere Model v.4 (CAM4) developed by the  National Center for Atmospheric Research
(NCAR).
For an in-depth technical description of CAM4, see \cite{Neal2010}. We configured the model 
to simulate an Earth-size aquaplanet (i.e.  globally ocean covered) in synchronous rotation 
(1:1 spin-orbit resonance) around M and late-K dwarfs. 
We ran the model at a horizontal resolution of 
$1.9^{\circ}\times2.5^{\circ}$ with $26$ vertical levels. The assumed atmosphere consists of 1 bar of N$_{2}$,
 1 ppm of $\co2$
and no $\ch4$, O$_{2}$, O$_{3}$, trace gases, or aerosols. We used a thermodynamic (slab) ocean model with 
a uniform depth of 50 m 
(a standard value for slab ocean models) and zero ocean heat transport.  The ocean surface  albedo is $0.06$
in the visible and 0.07 in the near-IR. We assumed zero obliquity and zero eccentricity,
 consistent with the assumption of
synchronous rotation; thus, there is no 
seasonal cycle.  Stellar spectra are assumed to be blackbody to facilitate comparison with
\cite{Yang2014a}.  
We use a model time step of 
1800 seconds.  Sub-grid-scale cloud parameters were taken as the default values for our given resolution and 
dynamical core selections.
Following a methodology similar to that of Yang et al. (2014), we moved planets closer to the star (by increasing the 
stellar flux) until the the model becomes unstable.  We used the last converged solution as a proxy for 
the inner edge of the HZ.

	\label{sec2}

	\section{Results}
	\label{results}
\subsection{Calculation of Correct Orbital Periods}

Following 
Kepler's third law, the orbital period $P$ is a function of  
orbital semi-major axis $a$ and the stellar mass $M_{\star}$:
\begin{eqnarray}
P \mathrm{(years)} &=& \left[\frac{a \mathrm{(AU)}^{3}}{M_{\star}/M_{\odot}}\right]^{1/2}
\end{eqnarray}
%Here, $P$ is in years and $d$ is in AU.
The stellar flux incident on a planet, $F_{P}$, compared to the flux on Earth ($F_{\oplus}$) can be written as:
\begin{eqnarray}
F_{P}/F_{\oplus} &=& \frac{L_{\star}/L_{\odot}}{ a^{2}}
\end{eqnarray}
where $L_{\star}/L_{\odot}$ is the luminosity of the star with respect to the bolometric luminosity of the Sun. 
Combining these two equations, the orbital 
period can be written  as:
\begin{eqnarray}
P (years) &=& \left[\left(\frac{L_{\star}/L_{\odot}}{F_{P}/F_{\oplus}}\right)^{3/4}\right] \left[{M_{\star}/M_{\odot}}  \right]^{-1/2}
%P (years) &=& \left[\frac{\left(\frac{L_{\star}/L_{\odot}}{F_{P}/F_{\oplus}}\right)^{3/2}}{M_{\star}/M_{\odot}}  \right]^{1/2}
\label{finalp}
\end{eqnarray}

This equation indicates that the orbital period of a planet in synchronous rotation can be calculated from the luminosity, mass and the incident flux,
and it cannot be chosen irrespective of the stellar type. This result is obvious, and not new, but it has significant 
implications for where the inner edge of the HZ is located. Moreover, for a star with a fixed effective radiating temperature, 
$T_{eff}$,
stellar luminosity
depends on the metallicity of  a star \citep[see Fig. 2a]{ChabBara2000, Dotter2008} because of
changes in the opacity of the stellar atmosphere. So the HZ also
varies as a function of stellar metallicity \citep{YoungP2012}. In this srudy, we used the
Dartmouth stellar evolution database \citep{Dotter2008}\footnote{\url{http://stellar.dartmouth.edu/models/ref.html}} 
to obtain $L_{\star}$ and $M_{\star}$ for both low ($[Fe/H] = -0.5$) and 
high ($[Fe/H] = 0.3$) metallicities, 
as a function of $T_{eff}$.
For convenience, we parameterized the relationship between $T_{eff}$, $M_{\star}$, and $L_{\star}$ and provided the
coefficients in Table \ref{table1}\footnote{\cite{Mann2015} have recently obtained a relation between
$T_{eff}$, radius of a star and metallicity for 183 nearby K7–M7 single stars. At the time this study 
was published, we had already begun our large suite of simulations. We will incorporate their relations
in a future study.}:

\begin{eqnarray}
y &=& a + bx + cx^{2} + dx^{3} + ex^{4} + fx^{5}
\label{lummass}
\end{eqnarray}
Here, $y$ can be either $M_{\star}/M_{\odot}$ or $L_{\star}/L_{\odot}$, and $x$ is $T_{eff}$. These relations apply only for
stars with $3300 \le T_{eff} \le 4500$K, as we are focused on low-mass stars which may have synchronously rotating planets near the
inner edge of their HZs. The lower limit of $3300$K is  based on the completeness of the Dartmouth model grid
for our chosen parameters.

\begin{threeparttable}[h!]
%\begin{center}
\caption{Coefficients to be used in Eq.(\ref{lummass}) to calculate stellar luminosity and mass from $T_{eff}$. }
\vspace{0.1 in}
\centering
\begin{tabular}{|c|c|c|c|c|c|c|c|c|}
\hline
[Fe/H]&$y$&$a$&$b$&$c$&$d$&$e$&$f$ \\
\hline
&$L_{\star}/L_{\odot}$&$-48.048$&$0.0645$&$-3.442 \mathrm{E}-5$&$9.098 \mathrm{E}-9$&$-1.192 \mathrm{E}-12$&$6.209 \mathrm{E}-17$\\
$-0.5$&&&&&&&\\
&$M_{\star}/M_{\odot}$&$874.787$&$-1.121$&$5.709 \mathrm{E}-4$&$-1.445 \mathrm{E}-7$&$1.821 \mathrm{E}-11$&$-9.138 \mathrm{E}-16$\\
\hline
&$L_{\star}/L_{\odot}$&$-27.044$&$0.033$&$-1.640 \mathrm{E}-5$&$4.036 \mathrm{E}-9$&$-4.961 \mathrm{E}-13$&$2.452 \mathrm{E}-17$\\
$0.3$&&&&&&&\\
&$M_{\star}/M_{\odot}$&$-406.344$&$0.480$&$-2.268 \mathrm{E}-4$&$5.342 \mathrm{E}-8$&$-6.271 \mathrm{E}-12$&$2.935 \mathrm{E}-16$\\

\hline
\end{tabular}
\label{table1}
\end{threeparttable}

One can calculate a range of orbital periods using Eq.(\ref{finalp}), substituting the values for $M_{\star}/M_{\odot}$ and
 $L_{\star}/L_{\odot}$ obtained from Eq.(\ref{lummass}), assuming a given $F_{P}/F_{\oplus}$. 
 
For a $3400$K star, 
using Eq.(\ref{lummass}), the corresponding stellar luminosity and mass are $0.0049 L_{\odot}$ and $0.189 M_{\odot}$, respectively,
for the low-metallicity case, $0.0204 L_{\odot}$ and $0.416 M_{\odot}$ for the high-metallicity case. 
A high metallicity star has a higher 
  mean atmospheric opacity, so for a certain stellar mass and age, a given optical depth is higher in the atmosphere and therefore at a lower temperature (lower Teff). To match the same Teff as a lower metallicity star (as is done here), the higher metallicity star (at a given age) must be more massive, increasing the overall stellar energy output and therefore Teff.
If one assumes  $P = 60$ days in Eq.({\ref{finalp}}), the corresponding stellar fluxes are $0.165 F_{\oplus}$
 for 
$[Fe/H] = -0.5$, and $0.406 F_{\oplus}$  for $[Fe/H] = 0.3$. 
These fluxes are much lower than the value used by \cite{Yang2014a} at the inner HZ; indeed, they are
substantially  lower than that of 
present day Earth.  Thus, a planet with $P=60$ days around a $T_{eff} = 3400$K  is more likely to have a frozen surface 
than to be at the precipice of a runaway.

\begin{figure}
\includegraphics[width=.85\textwidth]{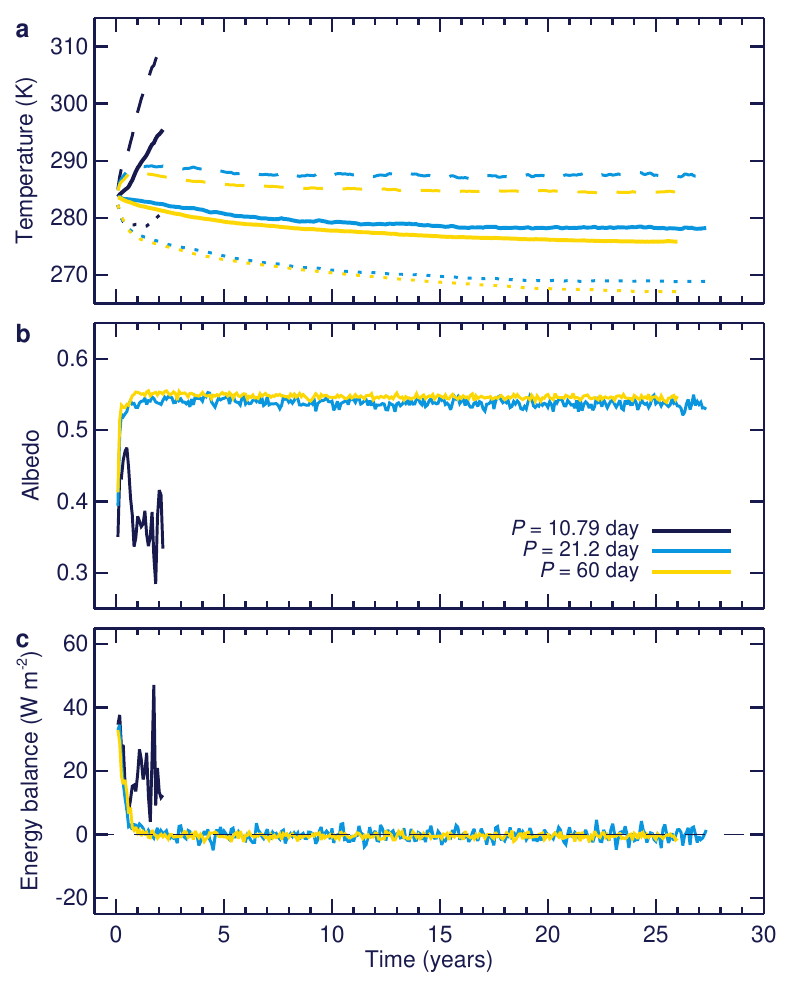}
\caption{Time series data from three simulations: $[Fe/H] = -0.5$ ($P= 10.79$ day), $[Fe/H] = 0.3$ ($P = 21.2$ day), and $P = 60$ day 
period.  Each simulation is irradiated by $2211$ W m$^{-2}$ of solar radiation. This value is the inner edge flux for a star with
$T_{eff} = 3400$K, as calculated by \cite{Yang2014a} for their slowly rotating case. 
Panel {\it a} shows the temporal evolution of the surface temperature, panel {\it b} shows the planetary albedo, and 
panel {\it c} shows the top-of-atmosphere energy balance (net solar in minus net longwave out).  In all figures solid curves show 
the global mean quantity.  Dashed and dotted curves in panel a show the dayside and nightside mean temperatures,
 respectively.}
\label{timeseries}
\end{figure}

  One can calculate a consistent orbital period from Eq.(\ref{finalp}), assuming an
inner HZ flux of $1.626 F_{\oplus}$ calculated from \cite{Yang2014a} for a $3400$K star. Then,
 the orbital period that the planet
{\it should} have is $10.79$ days for $[Fe/H] = -0.5$, and $21.21$ days for $[Fe/H] = 0.3$, both much shorter than the $60$-day
period assumed by \cite{Yang2014a}. We performed simulations at both of these orbital periods. For the 
$10.79$-day period,   the planet 
warms rapidly and no converged solution is found.
 By contrast, the $21.2$-day case and the (unphysical) $60$-day case remain climatologically stable 
(Fig.\ref{timeseries}). 
Both the 21.2-day and 60-day simulations exhibit similar climatological characteristics, 
with planetary albedos of $0.54$ and mean surface temperatures in the upper end of 270 K.
  While the $10.79$-day case is not quite a ``fast'' rotator like Earth,
  the increased rotation rate alters dynamics such that some of the 
substellar cloud deck  is advected to the anti-stellar side of the planet, lowering the planetary albedo, and causing the planet to become much warmer than in the slowly 
rotating cases.

For slower rotation rates ($P > 20$ days), thick substellar clouds produce a large planetary albedo, which stabilizes the climate 
against strong solar radiation. However, as $P$ approaches $ \sim 10$ days, the sub-stellar cloud decks begins to be smeared around 
the planet, resulting in a reduced albedo on the substellar hemisphere, and thus rapidly rising temperatures. The simulation with 
$P = 10.79$ days experiences numerical instability after only a few years, with a residual energy imbalance of $>10$ Wm$^{-2}$. 
While stable climates may lie beyond the numerical limits of our model, such atmospheres surely are excessively hot.

\begin{figure}
\includegraphics[width=.85\textwidth]{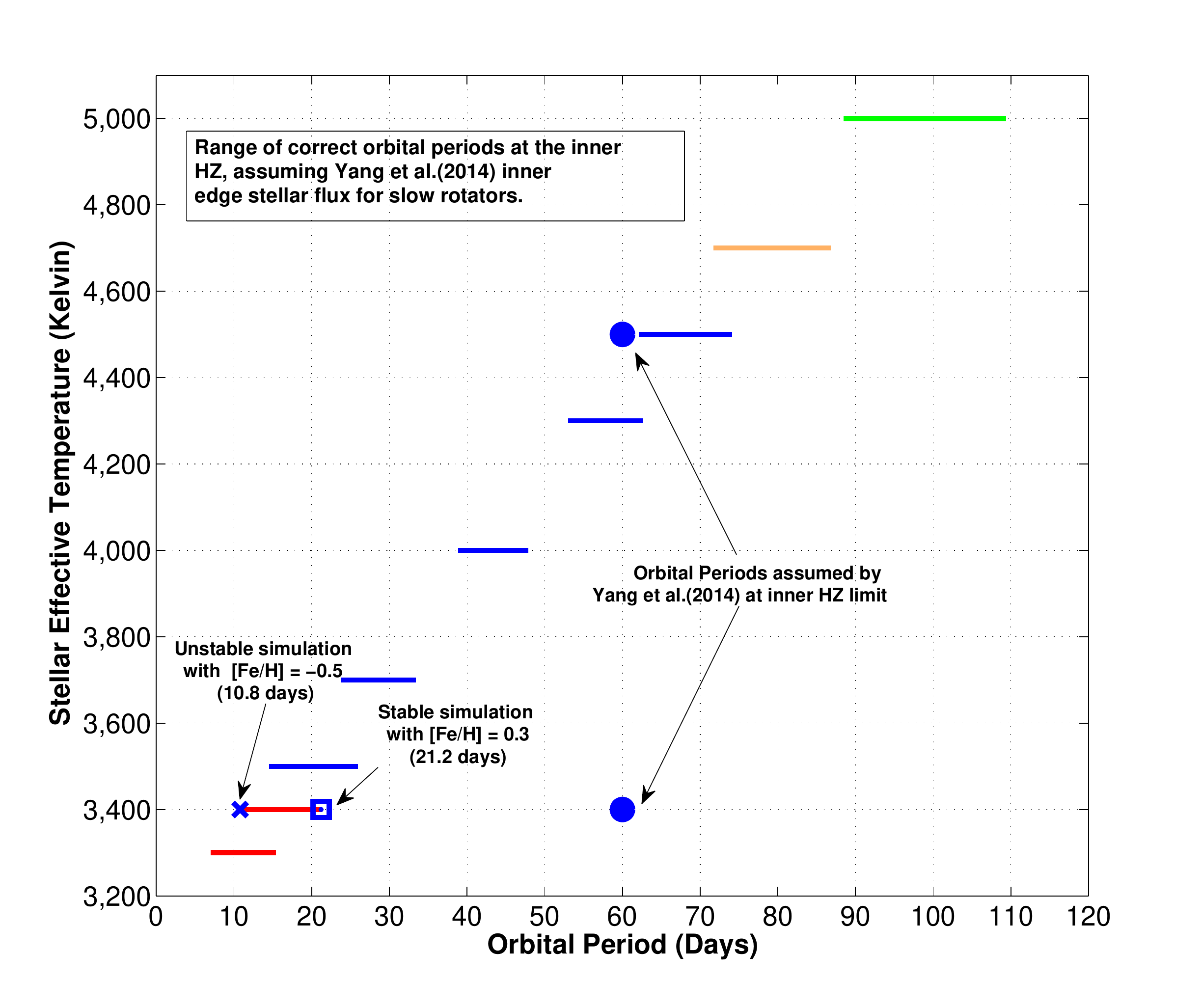}
\caption{Range of correct orbital periods, $P$, near the inner edge of the HZ for various stars (color horizontal lines),
 calculated using Eq.(\ref{finalp}), assuming an inner HZ flux for slow
rotators from \cite{Yang2014a}
for both high ([Fe/H] $= 0.3$) and low ([Fe/H] $= -0.5$) metallicity cases. Also shown are the $P = 60$ day periods
(blue filled circles) {\it assumed} by \cite{Yang2014a} at the inner edge of the HZ, which
 are inconsistent with 
Kepler's third law. Planets at the inner HZ limit are in 1:1 spin-orbit resonance. Using correct orbital/rotational 
periods for a $T_{eff} = 3400$K star, we find that the inner HZ flux from \cite{Yang2014a} warms a planet beyond the
stability limit of our 3-D model (blue cross), around a low [Fe/H] star.  }
\label{yangp}
\end{figure}

One can derive a range of orbital periods that is self-consistent with incident flux (Eq.(\ref{finalp})),
 using the luminosities and masses from Table \ref{table1} for different [Fe/H]. Fig.\ref{yangp} shows the 
{\it correct} orbital periods calculated using Eq.(\ref{finalp}) that one should use, if one assumes an inner HZ flux 
$F_{P}/F_{\oplus}$ from the 
slow-rotator equation of \cite{Yang2014a} for different stars. Also shown are the $P = 60$ day periods {\it assumed}
by \cite{Yang2014a} (blue filled circles), and the results from our own simulations for a $T_{eff} = 3400$K star
using the correct orbital periods ($10.8$ and $21.2$ days, blue cross and square, respectively) from Eq.(\ref{finalp})
 to illustrate the differences. The unstable $10.8$-day and the stable $21.2$-day solutions indicate that, for this 
$T_{eff} = 3400$K, the correct inner HZ limit should lie somewhere in between these two orbital periods.

To determine an inner HZ limit that is self-consistent with the calculated orbital periods, we followed the 
procedure described below for each star with $T_{eff} = 3300$K, $3700$K, $4000$K, $4500$K. The lower limit of 
$3300$K pushes across the edge of the Dartmouth model grid in some parameters, so we limit our simulations to 
values above this
$T_{eff}$. 
We first calculated the appropriate values of
$L_{\star}/L_{\odot}$ and $M_{\star}/M_{\odot}$ from Eq.(\ref{lummass}) for that $T_{eff}$, and then chose a 
range of $F_{P}$ (in Wm$^{-2}$) to calculate the corresponding orbital periods for these $F_{P}$. Then, we 
performed climate simulations at these $P$ and $F_{P}$ values using CAM, assuming synchronous rotation,
BB spectral energy distribution (SED) from the star, and zero 
planetary obliquity. We assumed, as did \cite{Yang2014a}, 
that the last converged solution represents the inner edge of the HZ. The SEDs used in our models are given in
Table \ref{sedtable}.

\begin{center}
\begin{threeparttable}[h!]
\caption{Blackbody spectral energy distribution of stars that were considered in this study, written in terms of
percentage of total flux in CAM wavelength bands.}
\vspace{0.1 in}
\centering
\begin{tabular}{|c|c|c|c|c|c|c|}
\hline
Band& $\lambda_{min}$ (micron) & $\lambda_{max}$ (micron) & $3300$K $\%$ flux & $3700$K $\%$ flux & $4000$K $\%$ flux & $4500$K $\%$ flux\\
\hline
1 & 0.200 & 0.245 &0.001913& 0.009296&0.02462&0.086219 \\
&&&&&&\\
2 & 0.245 & 0.265 &0.004174& 0.01659&0.03865&0.11277 \\
&&&&&&\\
3 & 0.265 & 0.275 &0.004139& 0.01497&0.03286&0.088338 \\
&&&&&&\\
4 & 0.275 & 0.285 &0.005468& 0.01865&0.03948&0.10097 \\
&&&&&&\\
5 & 0.285 & 0.295 &0.007856& 0.02528&0.05164&0.1257 \\
&&&&&&\\
6 & 0.295 & 0.305 &0.01225& 0.03724&0.07343&0.1703 \\
&&&&&&\\
7 & 0.305 & 0.350 &0.1158& 0.3045&0.5495&1.306 \\
&&&&&&\\
8 & 0.350 & 0.640 &8.3733& 12.902&16.913&22.73 \\
&&&&&&\\
9 & 0.640 & 0.700 &3.7251& 4.7689&5.5102&6.1537 \\
&&&&&&\\
10 & 0.700 & 5.000 &84.87& 79.63&75.524&67.853 \\
&&&&&&\\
11 & 2.630 & 2.860 &2.1099& 1.6653&1.4339&1.0786 \\
&&&&&&\\
12 & 4.160 & 4.550 &0.8054& 0.6082&0.5099&0.3701 \\
&&&&&&\\
\hline
\end{tabular}
\label{sedtable}
\end{threeparttable}
\end{center}

\subsection{Effect of Correct Orbital Periods on Climate}

In this section we discuss the large-scale atmospheric dynamics of
synchronously rotating planets. We examine simulations with a stellar
effective temperature of 3300 K at the correct orbital period of 9-day
which is self-consistent with the assumed incident flux, and for the 60d period, which is not self-consistent.   
As the planets are assumed to be synchronously rotating, such that the orbital period equals the period of rotation 
for the planet,
we focus on the dynamical
contribution toward the reduction in cloud cover caused by these differences
in rotation rate.

\begin{figure}
\includegraphics[width=.98\textwidth]{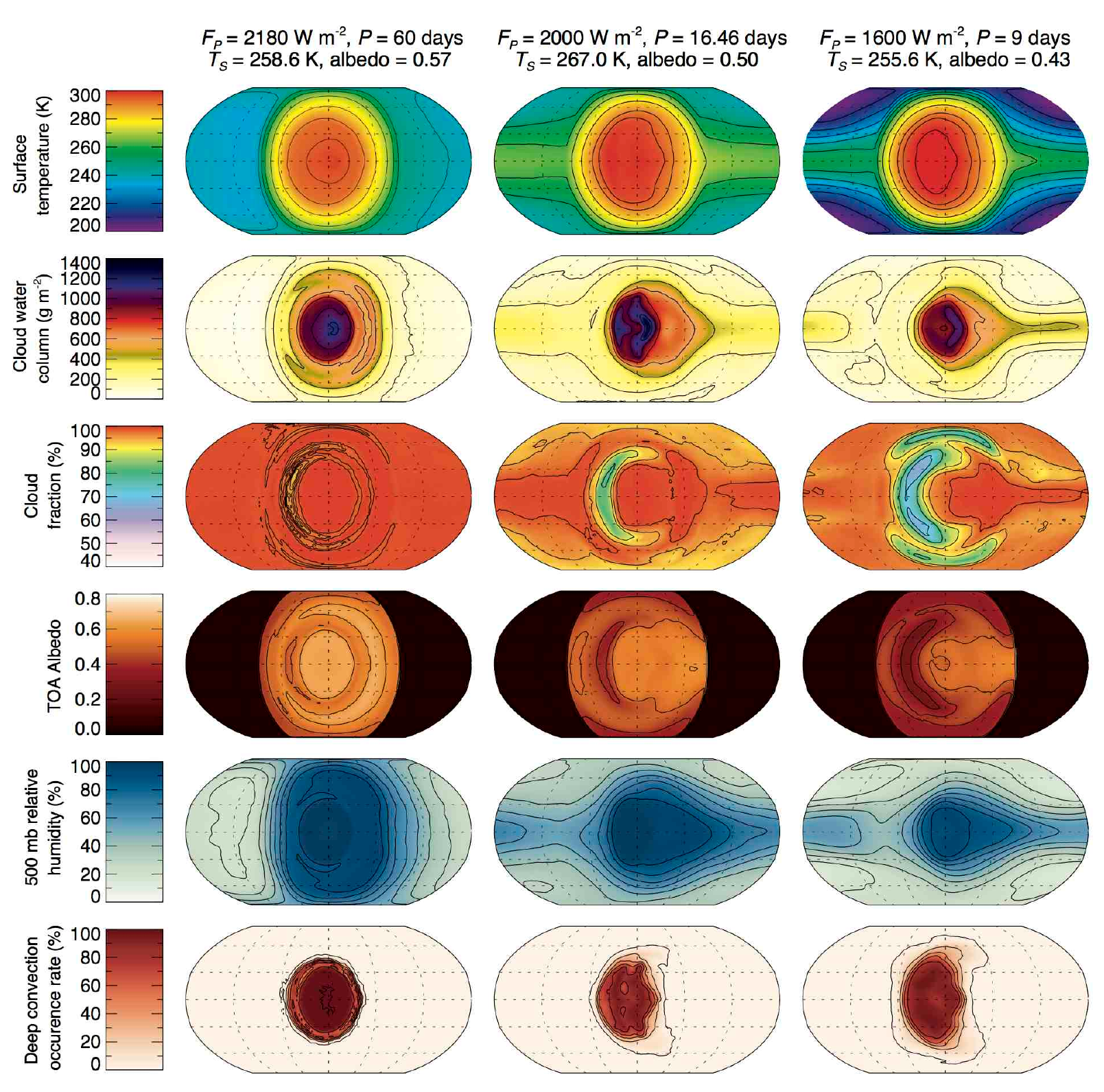}
\caption{From top to bottom rows: global mean surface temperature, column-integrated cloud water content, 
column-integrated cloud fraction, top-of-the atmosphere (TOA) albedo, 500 mb relative humidity, and deep 
convection occurrence rate.  These contour maps are shown for three different simulations, each with a 
different rotation rate around
a star with $T_{eff} = 3300$K. The first column assumes P = 60 days, following
Yang et al. (2014a). The remaining two column figures are the last converged solutions (i.e, planets at the inner HZ limit) for 
high and low stellar metallicities, whose Ps are calculated self-consistently with the incident flux (FP ) using Eq.(3). 
These results show that, at least for late M dwarfs, the final converged solution occurs at a lower incident flux,
 and so the 
inner HZ limit is not as close as predicted by Yang et al. (2014a).  
 A self-consistent calculation of P yields faster rotating planets, decreasing the substellar clouds and albedo, and thus increasing surface temperature.
}

\label{contour1}
\end{figure}

Fig. \ref{contour1} shows the latitude-longitude contour plots for three simulations of planets around a 
$T_{eff} =3300$K star, centered on the substellar point. The first column uses a $60$-day rotation period,
 as used by Yang et al. (2014a), to define the inner edge of the habitable zone for slow rotators. The second and third columns show 
contour plots for stable simulations using physically consistent orbital period and solar insolation relationships for both high 
and low stellar metallicity cases, respectively.  For each simulation, the incident flux, rotation rate, mean surface temperature and albedo are listed above each column.  While the incident stellar flux is different for each case, the resultant global mean surface temperatures are roughly similar.  The key difference between each case is the modification of the cloud albedo as a function of rotation rate.  Reducing the rotation rate from 60 to 9-days, decreases the top-of-atmosphere (TOA) albedo by 
$\sim 25\%$.

	Vigorous convection occurs over the substellar point in each of the three cases. 
Column-integrated cloud water and cloud fractions are expectedly large, as substellar convection lofts water into the free troposphere, creating high upper atmosphere relative humidities and thick convective water clouds, much like in 
Earth's tropics today. However, the important differences lie in the details. As the rotation rate of the planet increases from 60 to 16.46 to 9-days for high and low metallicity stars, respectively, the beginnings of a dynamical regime change become evident.  
This dynamical transition has previously been found by other studies for both highly-irradiated and cool planets of
different rotation rates \citep{HV2011, Carone2014, Showman2013a, KS2015}. While all cases maintain strong sub-stellar convection, gradually increasing the rotation produces a corresponding increase in the strength of the zonal winds (Fig 4., Fig 5).  Stronger upper level zonal winds begin to smear the clouds into a narrow equatorial band that can stretch around the entire planet.  The 500 mb relative humidity (and thus water vapor) closely mirrors the distribution of clouds.  The areal extent of clouds over the substellar hemisphere is reduced, most noticeably along the western flank of the sub-stellar point.  Increased upper level winds advect clouds eastward, away from their formation region above the sub-stellar point.  This reduces the overall fractional cloud cover on the sunlit side of the planet, decreasing the TOA albedo, and thereby warming the climate.  Thus, with a 9-day rotation rate, it takes 
$\sim 26\%$ less solar flux to maintain an equal global mean surface temperature compared to the 60-day rotation case.

	Note also that cloud fractions remain high ($\sim 100\%$) on the anti-stellar side of the planet, despite there 
being little to no convective activity, nor cloud water. On the anti-stellar side, the complete absence of solar surface heating, combined with substellar-to-antistellar atmospheric heat redistribution aloft, creates strong inversions, similar in nature to those that occur during the polar night on Earth. Here, a uniform, thin ice fog extends from the surface up to the inversion cap near 800 mb across the entire night side of the planet. Cloud fractions are near unity, but the clouds contain exceedingly little water and interact only very weakly with the thermal emission from the planet.

\begin{figure}
\begin{centering}
\includegraphics[angle=90,width=6.5in]{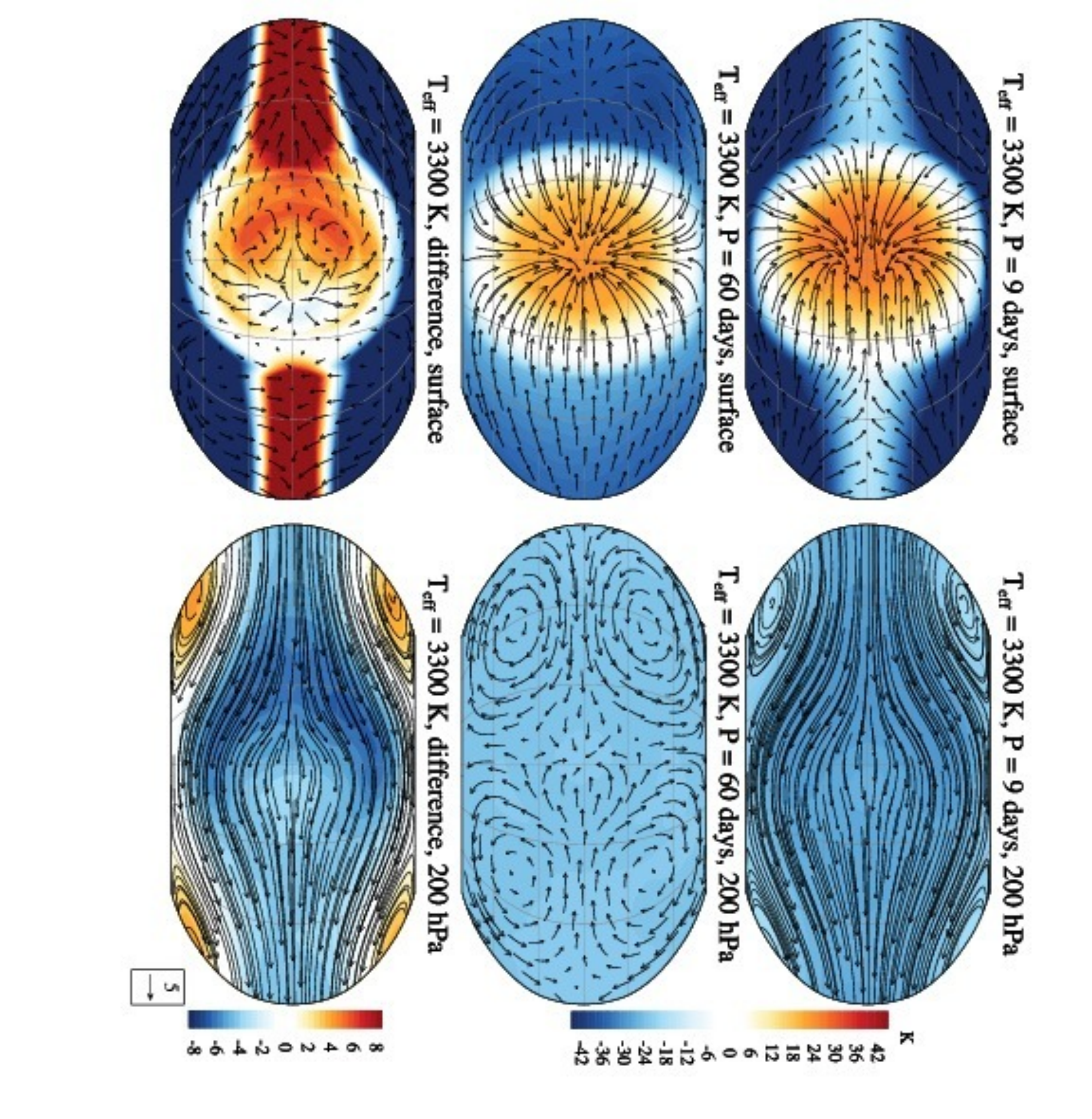}
\par\end{centering}

\protect\caption{Time average of temperature deviation from the freezing point of water
(shading) and horizontal wind (vectors) for T$_{eff}=3300$ K experiments at the
surface (left column) and upper atmosphere (right column). The top
row shows the 9d case, the middle row shows the 60d case, and the
bottom row shows the difference of the first row minus the second.
The substellar point is at the center of each panel. \label{fig:Temp}}
\end{figure}

Global average temperature and horizontal wind for $T_{eff} = 3300$K low metallicity,  9-day and 60-day
simulations are shown in Fig \ref{fig:Temp}. Surface
temperature and winds are shown in the left column of Fig \ref{fig:Temp}, with
inflow along the equator and from the poles into the substellar point
at the center. Upper atmosphere temperature and winds at the 200 hPa surface are shown
in the right-hand column of Fig \ref{fig:Temp}, with strong zonal westerly
flow aloft in the 9-day case and much weaker turbulent flow in the 60-day
case. Differences between these cases show up to 4 K of surface warming
at the substellar point and more than 10 K along the equator when
the rotation rate is increased. Warming of about 4 K also occurs
aloft in a gyre near the antistellar point. This warming occurs as
a result of a transition from the stabilizing cloud feedback described
by Yang et al. (2013) to a banded cloud structure that allows a larger
stellar flux incident on the surface. The
direction and magnitude of surface winds are similar between
the 9-day and 60-day cases, but the pattern of winds aloft differs significantly,
with the 9-day case displaying strong upper-level flow that generates
a banded cloud structure.

\begin{figure}
\begin{centering}
\includegraphics[width=6.5in]{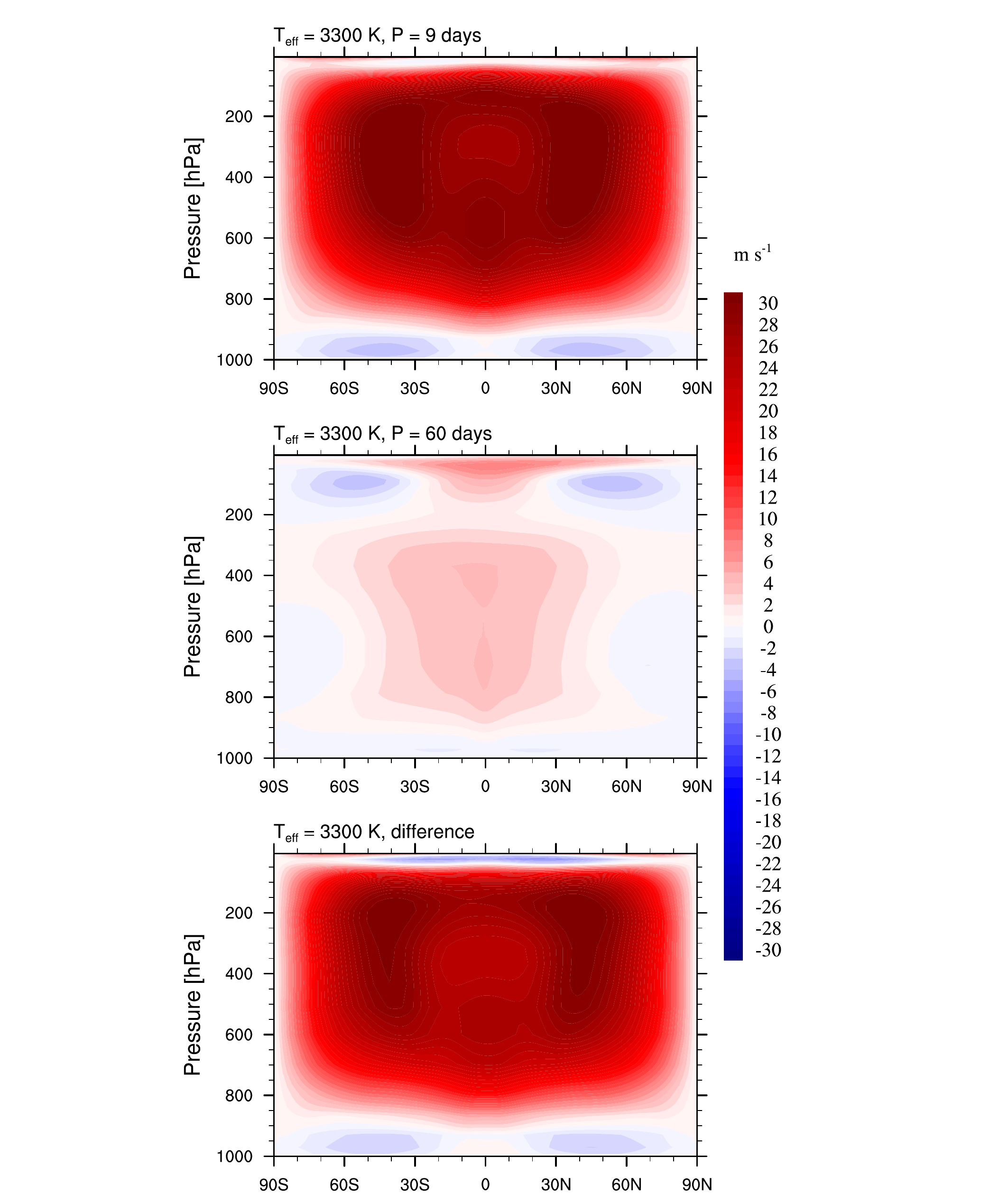}
\par\end{centering}

\protect\caption{Time average of mean zonal wind (shading) for T$_{eff}=3300$ K experiments.
The top row shows the 9d case, the middle row shows the 60d case,
and the bottom row shows the difference of the first row minus the
second. The substellar point is centered on the equator at the middle
of each panel. Positive (red) shading indicates westerly zonal wind,
and negative (blue) shading indicates easterly zonal wind.\label{fig:umean}}
\end{figure}

This shift in upper level flow is also apparent in the
time-averaged zonal mean zonal wind shown in Fig \ref{fig:umean}. The 9d
case shows strong westerly flow aloft at all latitudes, with two jets
prominent at midlatitudes and weaker easterlies along the surface.
By comparison, the 60d case shows weaker westerly flow aloft only
at tropical latitudes, with easterly jets in the upper atmosphere
toward the poles and extremely weak surface winds. This weaker flow
in the 60d case provides part of the physical process by which a stabilizing
cloud feedback can be maintained; however, the stronger flow when the rotation
rate is increased to 9d inhibits the maintenance of this cloud structure.

\begin{figure}
\begin{centering}
\includegraphics[width=6.5in]{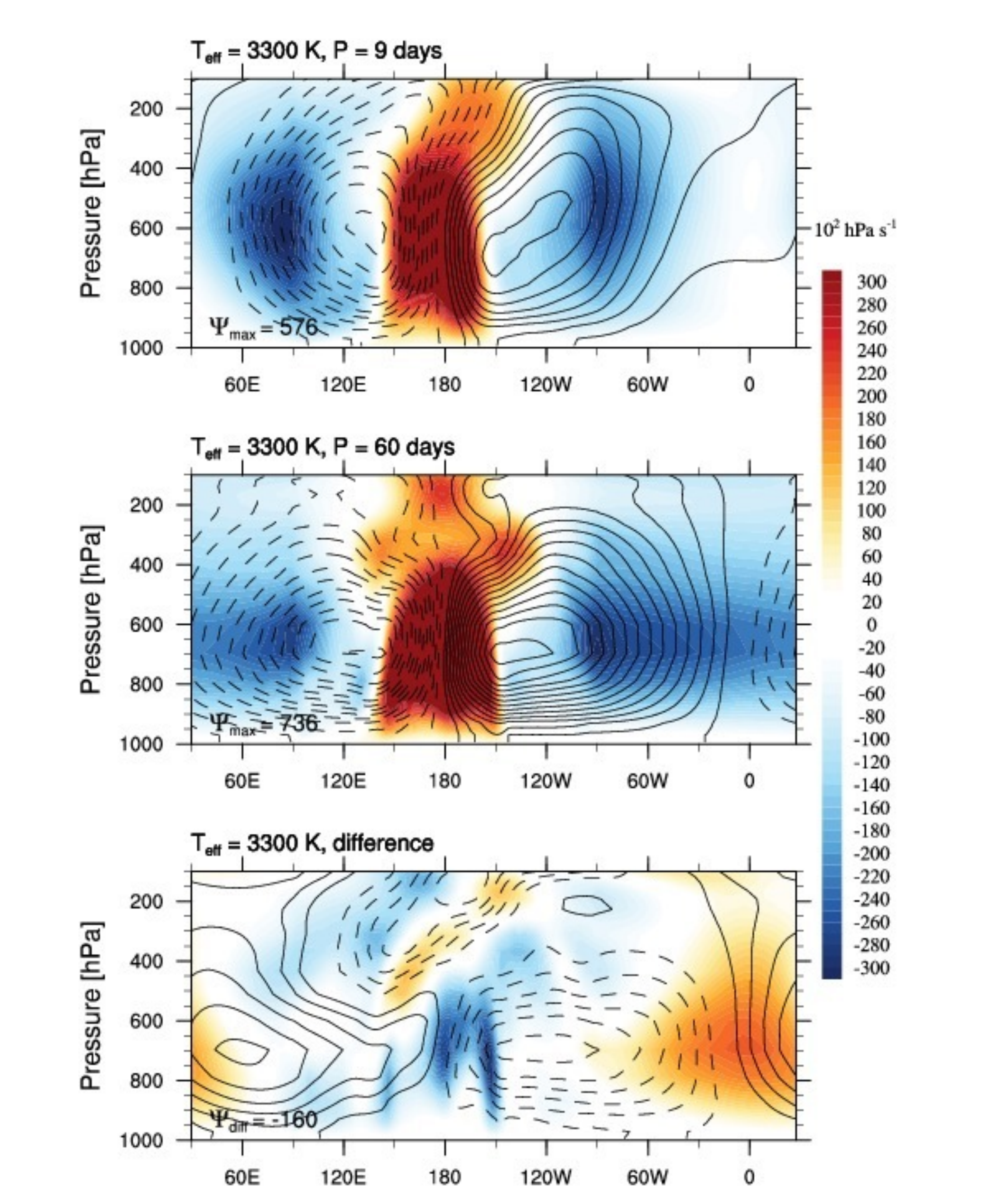}
\par\end{centering}

\protect\caption{Mean zonal circulation (line contours) and vertical wind (shading)
for the T$_{eff}=3300$ K experiments. The top row shows the 9d case, the middle
row shows the 60d case, and the bottom row shows the difference of
the first row minus the second. The substellar point is centered on
the prime meridian at the middle of each panel. The line contour interval
is $30\times10^{11}$kg s$^{-1}$ and the maximum streamfunction
$\Psi_{max}$ or difference $\Psi_{diff}$ is shown on each panel
in units of $10^{11}$kg s$^{-1}$. Solid contours indicate clockwise
circulation, and dashed contours indicate anti-clockwise circulation. Positive
(orange) shading indicates rising motion, and negative (blue) shading
indicates sinking motion. \label{fig:MZC}}
\end{figure}

The mean zonal circulation (MZC) and mean vertical wind for these
T3300 (i.e, T$_{eff} = 3300$ K simulations are shown in Fig \ref{fig:MZC}. Rising motion with
the MZC and upward vertical wind is apparent at the substellar point,
with regions of sinking in the vicinity of the antistellar point.
Both of these cases show that the MZC spans the full hemisphere from
substellar to antistellar point, which is expected because the orbital
periods of all our cases are greater than the critical period $P_{c}$
where a dynamical regime shift would occur (Edson et al. 2011; Carone
et al. 2014). The 60-day case shows a stronger zonal circulation than
the 9-day case, with greater rising motion at the substellar point
and sinking motion at the antistellar point. The symmetry of the MZC
and associated rising motion in the 60d case also suggests a symmetric
formation for clouds aloft, while the asymmetry observed in the 9d
case corresponds to the formation of banded clouds.

\begin{figure}
\begin{centering}
\includegraphics[width=6.5in]{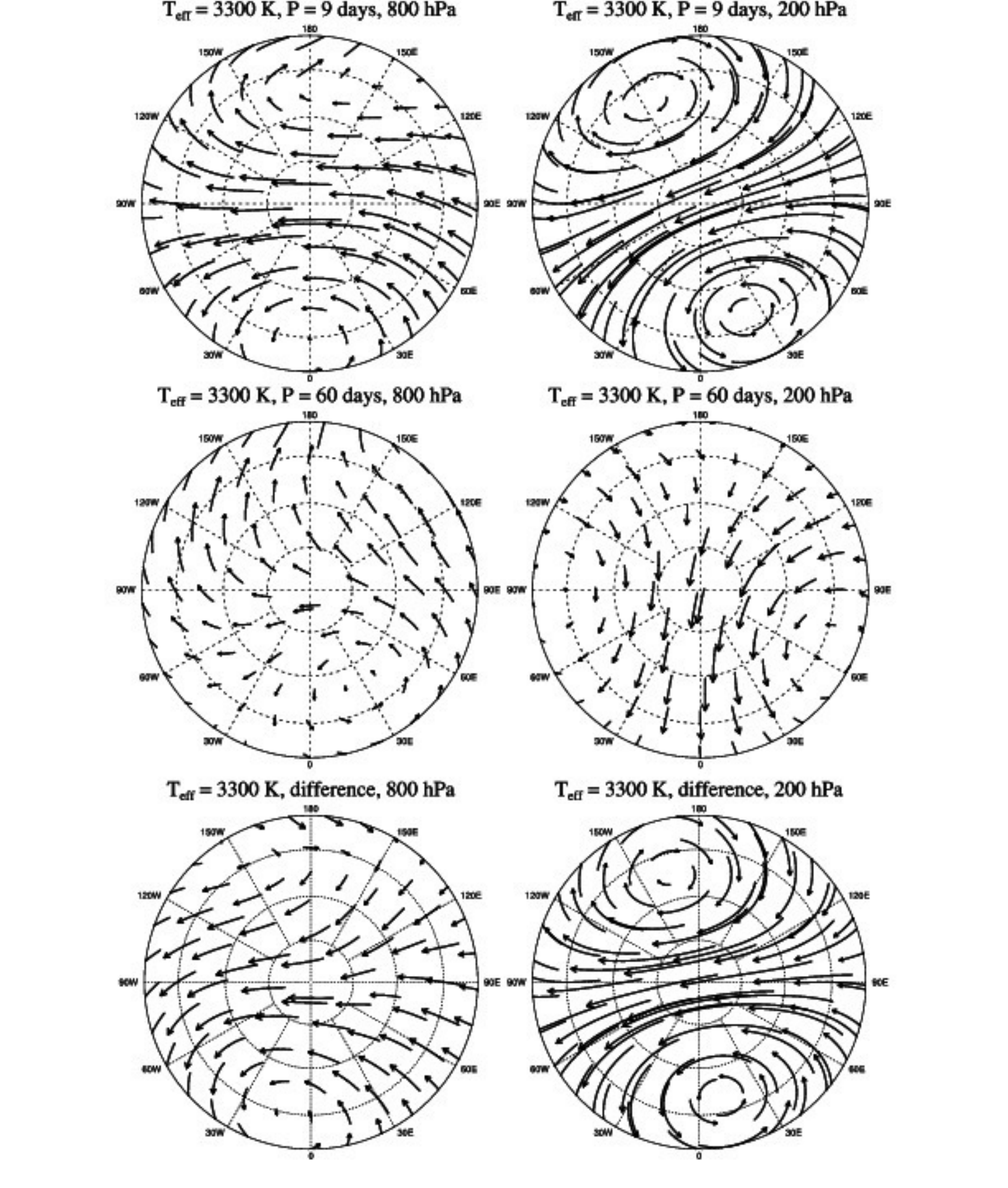}
\par\end{centering}

\protect\caption{Polar stereographic plots of horizontal wind (vectors) and eddy heat
flux (shading) for T$_{eff}=3300$ K experiments. The top row shows the 9d case,
the middle row shows the 60d case, and the bottom row shows the difference
of the first row minus the second. The substellar point is centered
on the equator and the prime meridian, which is shown at the top of
each panel. The plot shows the Northern hemisphere from $30^{\circ}$
to $90^{\circ}$ latitude, and the zonal mean is removed from all
zonal wind vectors.\label{fig:polarHeat}}
\end{figure}

The cross-polar circulation is shown in Fig. \ref{fig:polarHeat}.
This circulation pattern is obtained by subtracting the zonal mean
from all zonal wind vectors (Joshi1997; Haqq-Misra and Kopparapu
2015), which helps to illustrate the complex three-dimensional patterns
that form the general circulation of synchronous rotators. Both cases
show flow across the pole at both the 800 hPa and 200 hPa surfaces,
although the direction of this flow varies with each experiment. Both
cases also display vortices near $0^{\circ}$ to $30^{\circ}$ longitude
that contribute to a unique circulation pattern not easily described
by the MZC or the mean meridional circulation. This cross-polar circulation
is stronger in the 9-day case, both an aloft and along the surface, while
the 60d case shows a relatively similar flow pattern but with a reduced
magnitude. The increase in cross-polar flow when rotation rate changes
from 60d to 9d also provides a mechanism that would disrupt the formation
of a stabilizing cloud feedback above the substellar point.


\subsection{The Inner HZ Limit with Correct Orbital Periods}

\begin{figure}[!hbp|t]
\subfigure[]{
\includegraphics[width=.53\textwidth]{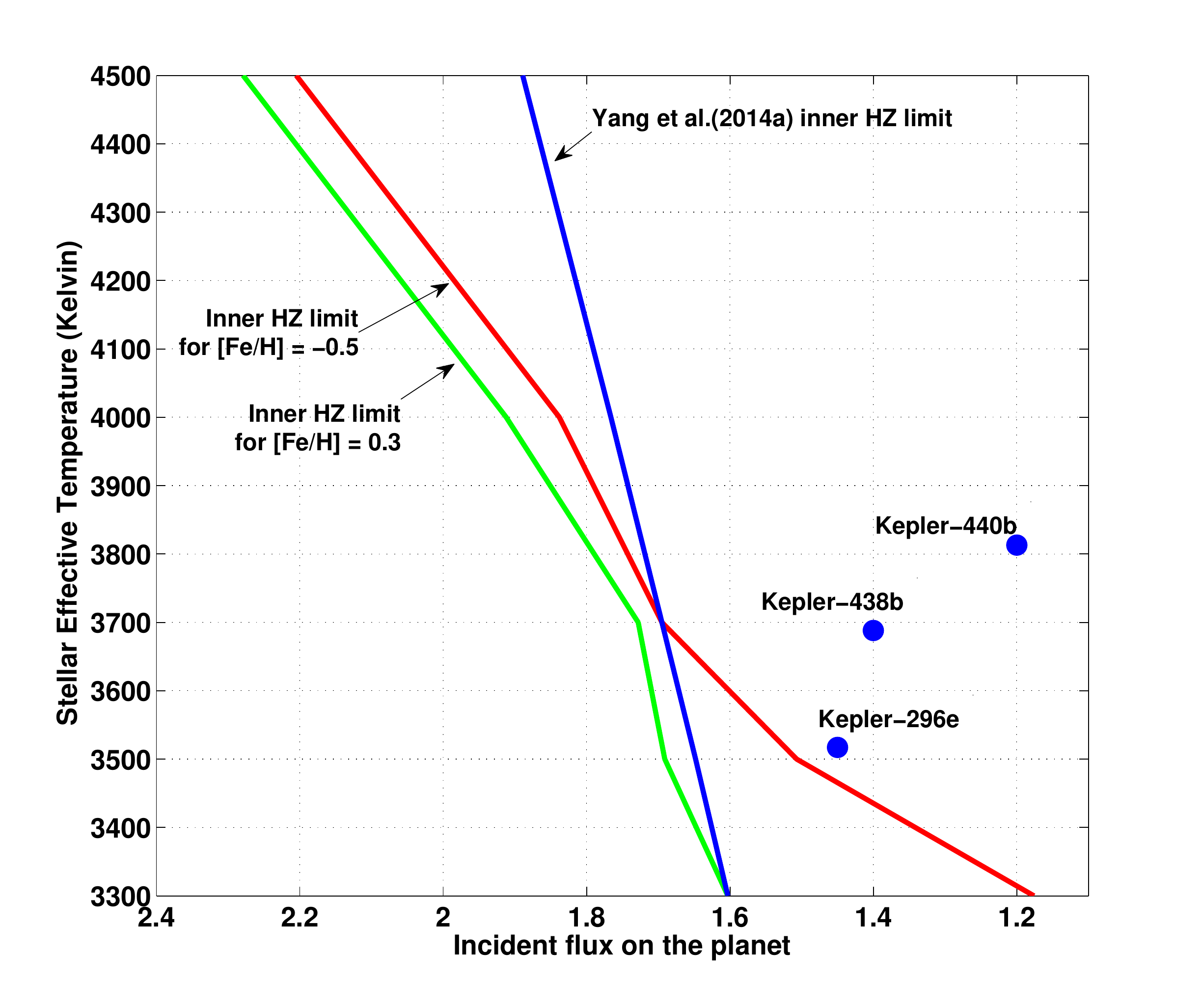}
}
\subfigure[]{
\includegraphics[width=.53\textwidth]{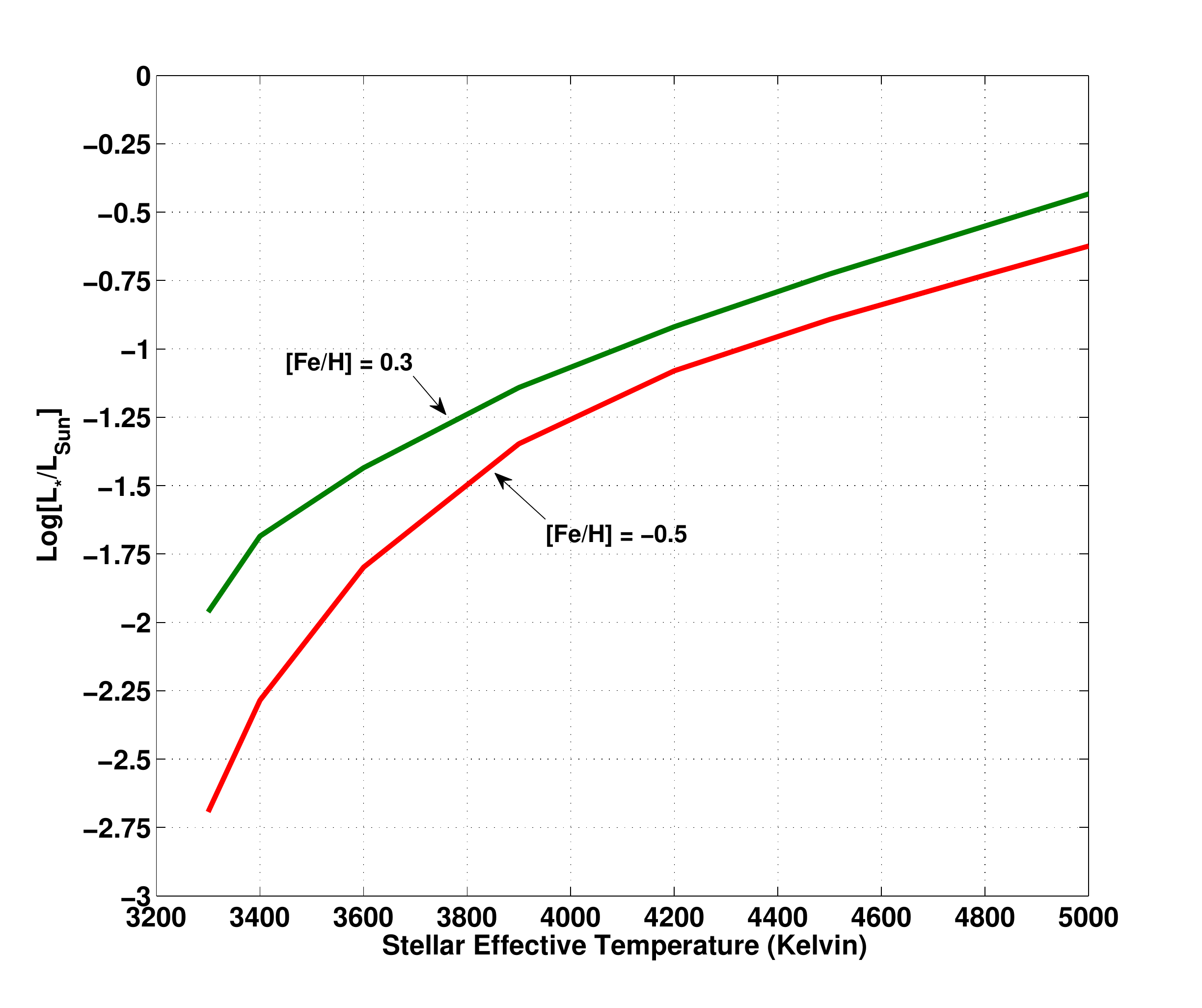}
\label{lumteff}
}
\caption{{\it Panel a:} Estimates of the inner edge of the HZ for low-mass stars from this study for [Fe/H] $=-0.5$ (red) and $0.3$ (green),
 compared with \cite{Yang2014a} (blue). {\it Panel b:} Dependence of the stellar luminosity on [Fe/H] \cite{Dotter2008}.
Synchronously rotating planets around lower luminosity stars (red) have faster orbital (and rotational) periods, and hence
have a tendency to develop banded cloud formations. This reduces the planetary albedo due to reduction of thick clouds
at the substellar point. Therefore, the inner HZ limit is at a lower stellar flux for these low-metallicity, low-luminosity
stars. }
%Since the orbital period depends on stellar luminosity, and e on the incident flux on the planet, 
%the large difference in luminosity for different metallicities shown in {\it panel b} translates to large differences in
%the incident stellar flux.}
\label{newhz}
\end{figure}

Fig. \ref{newhz} shows our new inner HZ boundary for stars with $3300$K $\le T_{eff} \le 4500$K. We show the
 limits for both
low (red) and high (green) [Fe/H] cases. The red curve is farther away from the star because, for a given stellar 
$T_{eff}$, the low-metallicity star is brighter (see discussion below). For comparison, we also plotted the inner HZ limit from \cite{Yang2014a}
 (blue) for their 60-day rotator case. These results indicate that the inner edge of the HZ is 
closer to the star than predicted by \cite{Yang2014a} for $3700$K $\le T_{eff} \le 4500$K. This is true for both the high
 and low [Fe/H] cases, though the difference between these two [Fe/H] is minimal. For stars with $T_{eff} \le 3700$K,  
although the low and high [Fe/H] cases diverge dramatically from each other, with the high [Fe/H] case (green) staying closer to the 
\cite{Yang2014a} inner edge limit (blue). 
%Our new definitions for the inner edge of the HZ
% indeed has a significant impact on the number of terrestrial size planets
%in the HZ. To illustrate this point, 
We have also shown several currently confirmed exoplanets on this plot.
%\footnote{We should note here that \cite{Yang2014a} also consider rapidly rotating planets
%and calculate the inner edge of the HZ. The shortest orbital period that remains stable in our simulations is 
%$\approx 9$ days. So we compare our results with the slow rotator case of \cite{Tang2014a}. }

The reason for this large divergence between the two stellar metallicity cases can be traced to the differences in the stars'
luminosities and planetary rotation rates. Fig. \ref{lumteff} illustrates this point. The difference between $L_{\star}/L_{\odot}$ for 
[Fe/H] = $0.3$ (blue) and $-0.5$ (red) is largest at lower $T_{eff}$, and that is precisely where we see the 
divergence in our
inner HZ result. The difference in log$[L_{\star}/L_{\odot}]$ for the two metallicities is $\approx 30 \%$ 
at $T_{eff} = 3300$K. 
%Since $F_{P}/F_{\oplus}$
%scales linearly with luminosity, this translates also to $\approx 30\%$ in the incident fluxes on the planet. 
%An additional effect that also contributes to this difference in the fluxes. 
To receive the same amount of 
incident $F_{P}$ around a low luminosity star, a planet that is in synchronous rotation needs to be at a shorter $P$ 
(fast rotator), according to Eq.(\ref{finalp}), compared to a high luminosity one. As discussed before, 
fast rotating planets have stronger Coriolis force, which tends to create more banded cloud formations 
that reduce the
planetary albedo.  Furthermore, stars with
lower $T_{eff}$ output more flux in the near-infrared (IR) part of the SED. As terrestrial planets at the
inner edge of the HZ are expected to have increasing amounts of water vapor in their  atmospheres, and as $\h2o$ is a good near-IR absorber, a small increment
in the stellar flux around these low $T_{eff}$ stars would push the planet close to the runaway greenhouse limit.
Hence, the inner edge of the HZ is further away around these cool stars.

	For stellar effective temperatures greater than 3700 K, we find that the inner edge of the 
HZ occurs at stellar fluxes up to 20\% larger than shown by \cite{Yang2014a}.  This result was unexpected, 
as in this regime orbital periods in the HZ remain firmly within the slow-rotating regime, and 
there is little difference between our model and theirs in the mean state of the climate.  In fact, for 
$T_{eff} = 4500$ K,  we find self-consistent orbital periods at the inner edge of the HZ of 
55.31 and 64.1 days for low- and high-metallicity stars, respectively, bracketing the 60-day value used in
\cite{Yang2014a}.  
	However, the differences can be traced to 
subtle differences in the 
model configuration.  \cite{Yang2014a} used CAM3, while here we use CAM4 with an improved numerical algorithm 
for deep convection.  With this patch, CAM4 is able to remain numerically stable at hotter 
temperatures, while keeping a relatively long model timestep (1800 seconds).  To better understand this difference, we repeated 
simulations for 4500 K stellar effective temperature star using CAM3
without any numerical improvement to the deep convection scheme, but with an exceedingly short model 
timestep (100 seconds) to improve numerical stability.  When we do this, we find similar results to those found with 
CAM4 using a longer timestep. Thus, there is no 
actual conflict between the results shown here and those shown in \cite{Yang2014a} for $T_{eff} > 3700$K; 
rather, it is an issue
 of model configuration and numerical stability. This point highlights why model intercomparisons are of the 
utmost importance.

\subsection{The nature of the last converged solution}
Computational and numerical limitations prevent current Earth-derivative 3D climate models from exploring the full 
runaway greenhouse process, through surface temperatures beyond the critical point and the vaporization of the 
entire oceans.  This has led to the so-called ``last converged solution'' criteria being used as a proxy for the 
inner edge of the HZ (Wolf and Toon, 2014; Yang et al. 2014).  While at first 
glance this criteria appears unsatisfying, the last converged solution may nonetheless be diagnostic of an 
imminent climatic state transition.  Using CAM4 to study Earth (a rapid rotator), Wolf and Toon (2014) first found 
the last converged solution to have a global mean surface temperature of  $\sim 313$ K.  However, in a subsequent 
study, using the same model but with improved numerics, Wolf and Toon (2015) showed that their previous assumed 
last converged solution was actually hovering just below the threshold of an abrupt transition into a hotter 
climate state.

\begin{figure}[!hbp|t]
\begin{center}
\includegraphics[width=.90\textwidth]{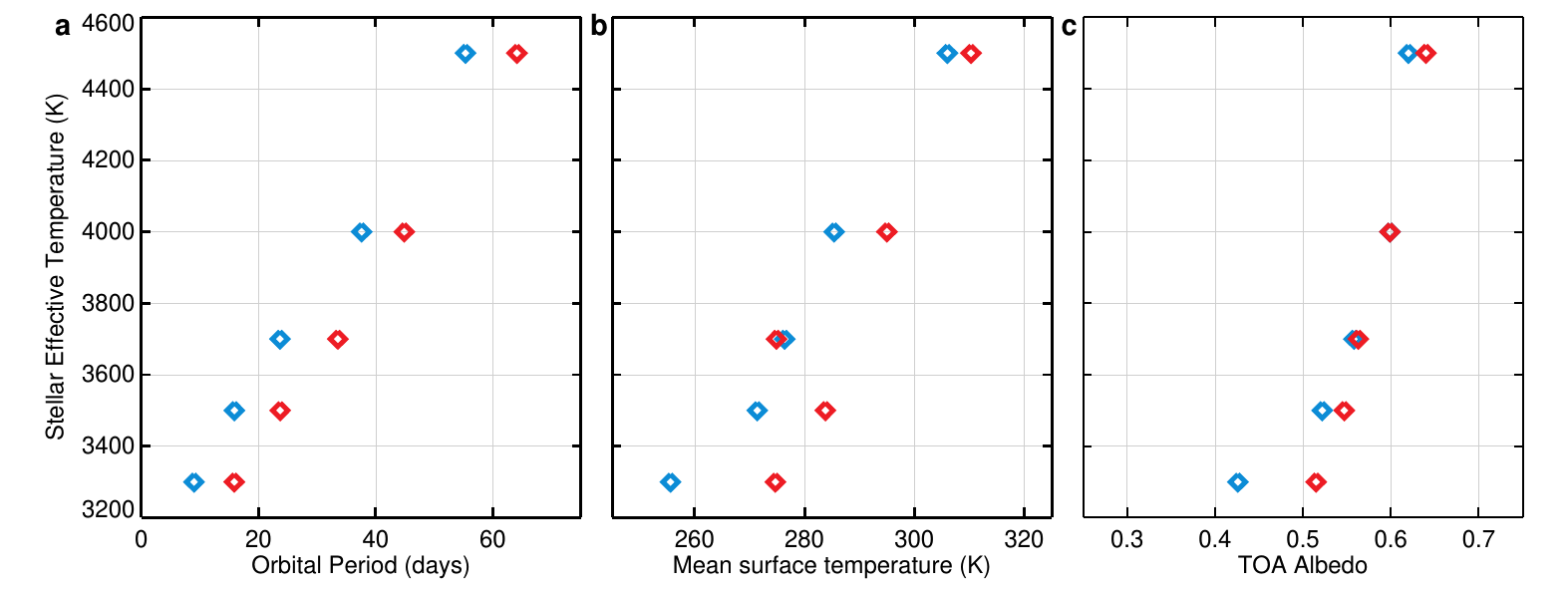}
\caption{Here we show the rotation rate (a), mean surface temperature (b) and top-of-atmosphere albedo (c) for 
the last converged solutions found in this study.  Blue diamonds are for low-metallicity stars, and red diamonds are 
for high-metallicity stars.  Variability about the trend is due to the coarseness of our solar constant interval 
(50 or 100  W m$^{-2}$, depending on $T_{eff}$) across our simulation sets.   The temperature of the 
last converged solution 
is much lower around redder stars, due to the stronger interaction of the SED with the near-IR water vapor bands.  
The trend in the TOA albedo is dominated by the effect of the orbital period, which controls thickness of the 
substellar cloud deck.}
\label{lastconv}
\end{center}
\end{figure}

In Fig. \ref{lastconv}, we summarize the global mean climates of our last converged solutions as a function of the 
stellar effective temperature.  There is a definite trend in the global mean surface temperature of the last 
converged solution, with the last stable temperatures being significantly lower around 3300 K stars compared to 
4500 K stars.  This trend is caused by the interaction of different spectral energy distributions (SEDs)
 with atmospheric water vapor. 
 Changes to the top of the atmosphere (TOA) albedo also display a trend, in this case linked most tightly to the rotation rate of the 
planet due to the dynamical modulation of the cloud fields, as is discussed in section 3.2.  SED does affect 
atmospheric scattering, with planets around bluer stars scattering more of the incident sunlight than those 
around redder stars; however this effect is relatively small at the atmospheric temperatures and pressures we 
consider in this study.  Here, the TOA albedo is dominated by the clouds rather than by atmospheric scattering.   
As the rotation rate (Fig. \ref{lastconv}) falls to $ \sim 10$ days, a steep reduction in the TOA albedo is observed.  

\begin{figure}[!hbp|t]
\begin{center}
\includegraphics[width=.90\textwidth]{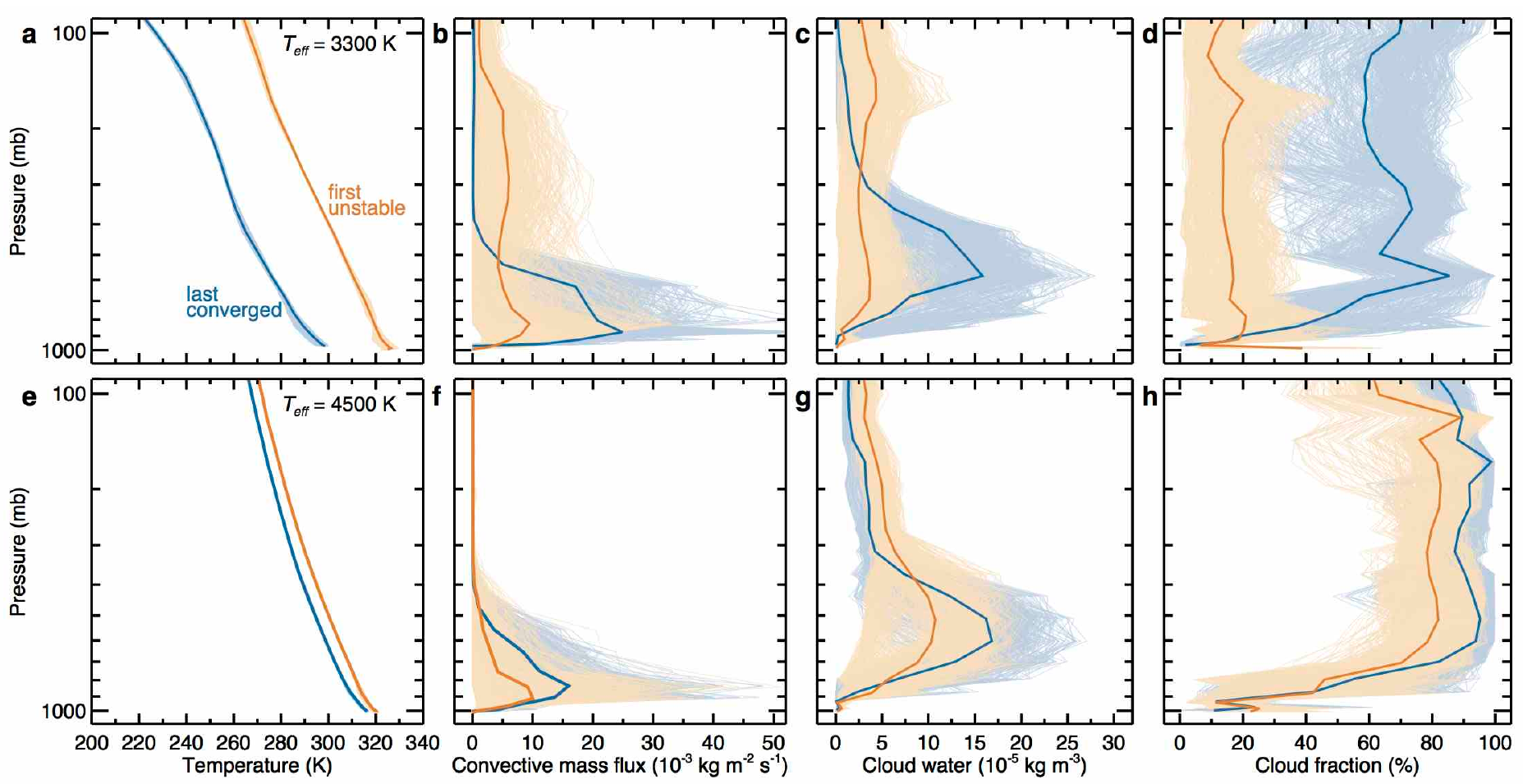}
\caption{This plot shows vertical profiles of temperature (a, e), upward convective mass flux (b,f), grid-box averaged 
cloud water content (c,g), and cloud fraction (d,h) for the substellar region of the planet; (the elliptical 
region bounded by $\pm 30^{\circ}$ latitude and longitude from the substellar point).  The top row is for planets around 
high-metallicity 3300 K stars, and the bottom row is for planets around high-metallicity 4500 K stars.  Blue lines are 
for the last converged solution, red lines are for the first unstable simulation.  The bolded lines are the mean of 
the substellar region.  The light colored lines are from each individual profile within the substellar region.  While 
the bifurcation is sharper around the 3300 K star, in both cases the first unstable simulation is characterized by a 
lessening of the lapse rate in the boundary layer and a reduction in substellar convection and thus in clouds.}
\label{substellar}
\end{center}
\end{figure}

To gain more insight into the problem, we must also consider what happens immediately beyond the last converged solution: 
the so-called ``first unstable'' simulation.  The first unstable simulation is defined as that with the smallest value for the 
solar insolation where the simulation fails due to numerical instability. Beyond the last converged solution, our test planet is moved 
closer to the star, and thus towards higher stellar fluxes and slightly shorter rotational periods.  Here, dynamical 
differences due to the change in rotation rate are not important, as the rotation rate only changes by several 
percent between the  last converged and the first unstable simulations.  The key driver is the input of solar energy.  
In  Fig. \ref{substellar} we show vertical profiles of temperature, convection, and clouds for the substellar 
region of the planet for the last converged and first unstable simulations around $T_{eff}$  = 3300 K and 4500 K high-metallicity 
stars.  Note, we show only profiles (and the mean) from the substellar region (the region forming an ellipse bounded 
by $\pm 30^{\circ}$ latitude and longitude from the substellar point), as this is where convection and clouds most 
strongly influence the energy budget of slow-rotating aquaplanets.  While converged simulations of course reach 
equilibrium conditions, unstable simulations fail well before reaching equilibrium.  For the 
first unstable simulations shown in Fig. \ref{substellar}, we take results averaged over the last 30 days of 
simulations immediately before the point of failure.

 Increased solar input drives the water vapor greenhouse feedback.  Increased atmospheric water vapor leads to 
increased solar absorption aloft, and thus reduces the mean lapse rate, noticeably below about 800 mb in 
Fig. \ref{substellar}.  This causes the substellar atmosphere to trend towards convective stability, reducing 
convection emanating from the substellar boundary layer (Figure 10b,f) and thus suppressing the formation of the 
ubiquitous substellar cloud deck (Figure 10c,d,g,h).  A similar mechanism for radiatively driven convective stabilization
 of the lower atmosphere was proposed for rapidly rotating planets (WP2013; Wolf \& Toon, 2015).
  Strong solar absorption in the near-IR water wapor bands can stabilize the lower atmosphere and prevent boundary layer 
convection from occurring.  However, while rapidly rotating planets may maintain climatological stability beyond this 
transition, the transition appears catastrophic for strongly irradiated synchronous rotators.  At the last converged 
solution, the substellar cloud deck protects synchronous rotators from incendiary warming under immense stellar fluxes.  Thus, even a small dent in the substellar cloud shield lets in a tremendous amount of solar radiation, strongly 
destabilizing climate. 
Note that this bifurcation about last converged and first unstable simulations is also illustrated in Fig. 1.

While the triggering of climatic instabilities in the model is caused by increases to the total stellar flux, the trend 
in the temperatures of the last converged solution are traceable to differences in stellar effective temperature and 
thus the SED.  The last stable temperature is much lower around lower $T_{eff}$ stars due to the fact that redder SED 
interacts much more strongly with near-IR water vapor absorption bands.  Thus, strong solar heating and subsequent 
convective stabilization around the substellar point can occur with less water vapor in the atmosphere and thus at 
lower mean surface temperatures around redder stars compared with bluer stars.  The bifurcation between the last 
converged and first unstable simulations is very pronounced around 3300 K stars (Fig. \ref{substellar}).  However, the
 apparent bifurcation becomes more subtle as one moves towards higher stellar effective temperatures.  Nonetheless, 
even for the 4500 K case, one can still clearly see that the first unstable simulation has less substellar convection 
and clouds than the last converged solution.  This result agrees with the assertion of Kopparapu et al. (2013), that 
planets around low mass stars may skip over the moist greenhouse phase completely and instead move straight to the 
runaway phase, owing to stronger interactions with water vapor near-IR bands.   Thus, planets around bluer stars may 
have additional stable climate states near the inner edge boundary.

\subsection{Estimating water-loss to space}
In section 3.3 we established new limits to the inner edge of the HZ based on the last converged solutions for 
stable climates.  This criterion is intentionally taken to compare with inner HZ results for slow rotators in 
Yang et al. (2014).  In section 3.4 we argue that the last converged solutions lie just below the threshold of 
an abrupt climatic transition towards hotter states, whereupon the substellar cloud deck weakens considerably 
and strong solar radiation then heats the planet.  However, we have not yet considered the potential for 
significant water-loss from our modeled atmospheres.  Notably, water-loss to space from a moist greenhouse is 
believed to be the more conservative estimate for inner edge of the HZ, and essentially a limit to habitability, 
occurring at a lower stellar flux (and thus further from the star) than does a runaway greenhouse.

\begin{figure}[!hbp|t]
\begin{center}
\includegraphics[width=.90\textwidth]{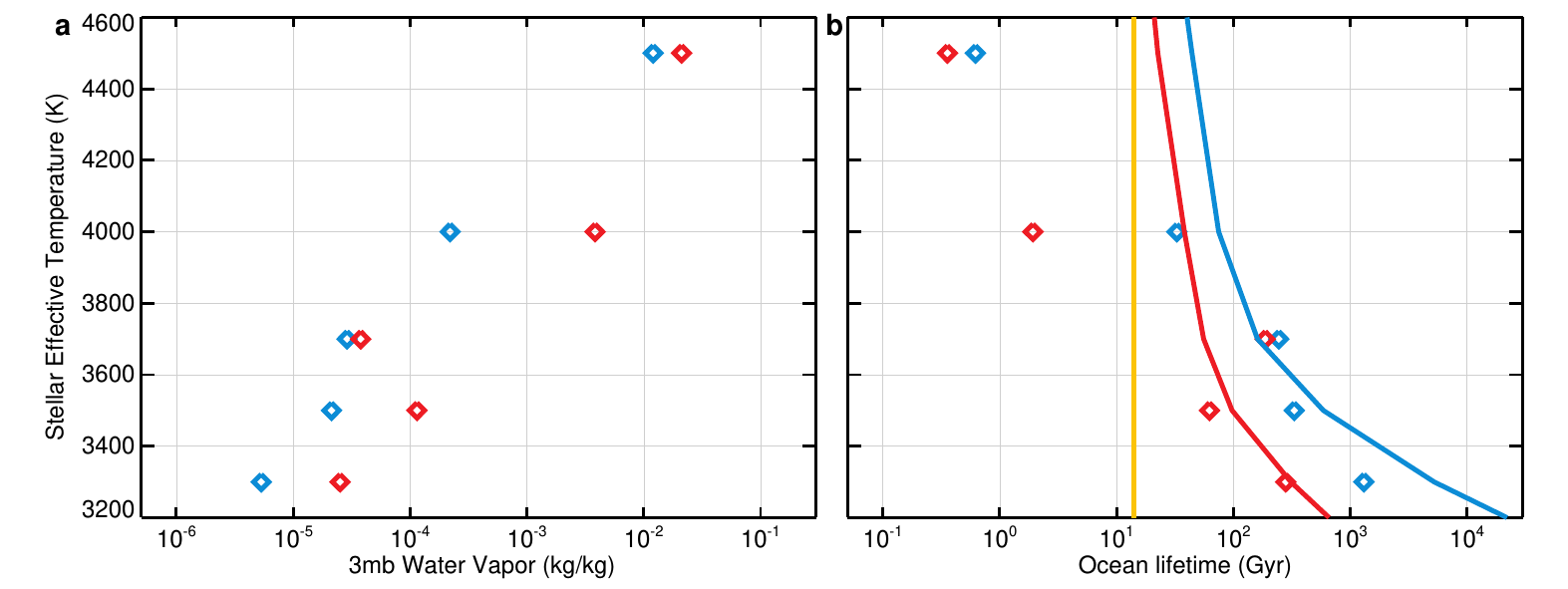}
\caption{The model top (3 mb) water vapor volume mixing ratio (a) and the lifetime of an Earth ocean against 
diffusion limited water-loss (b).  Blue and red diamonds are from the last converged solutions around 
low-metallicity stars high-metallicity stars respectively.  The solid blue and red lines in panel (b) are the 
total main sequence lifetime for low-metallicity and high-metallicity stars respectively.  The vertical yellow 
line in panel (b) is the current age of the universe. }
\label{waterloss}
\end{center}
\end{figure}

At first glance, Fig. \ref{lastconv} appears to indicate that our model atmospheres located nearest the inner edge of the HZ are probably too cold to experience significant water-loss to space, all being below ~ 310 K.  Note that the canonical value for the mean surface temperature marking the moist greenhouse water-loss limit is 340 K in 1D models (Kasting et al. 1993; Kopparapu et al. 2013), and was recently found be to ~350 K using CAM4 to simulate Earth under high stellar fluxes (Wolf \& Toon, 2015). In Fig. \ref{waterloss} we show the model top (i.e. 3 mb) water vapor mixing ratio and the lifetime of the ocean against water-loss, assuming an initial water inventory equal to 1 Earth ocean ($1.4 \times 10^{24}$g H$_{2}$O). We assume diffusion limited escape from the upper most model layer, following Hunten (1973).  The solid red and solid blue lines in the Fig. 11b are the estimated main sequence lifetimes for high and low metallicity stars respectively.   Interestingly, for nearly all our simulations that define the IHZ, even those with seemingly little water vapor in their upper atmospheres, the lifetime to lose an Earth ocean is approximately equal to or shorter than the main sequence lifetime of the host star.  Of course, for M stars this timescale remains much larger than the current age of the universe, and thus is of little practical use for determining habitability.  

However, synchronous rotators around late-K stars have significant water vapor near the model top and thus can have rapid water-loss rates.   Our last converged solutions around stars with $T_{eff} = 4500$K can lose an Earth ocean in less than 1 Gyr (red and blue diamonds in Fig. 11b, left-top).  Thus these atmospheres may not be habitable under the more stringent water-loss criteria.   The inner edge of the HZ defined by water-loss is thus likely at a lower stellar flux for late-K stars, compared with that illustrated in Fig 8. These atmospheres undergo rapid water-loss at a surprisingly low global mean surface temperature, between 295 K and 310 K.   We postulate that strong substellar convection may be more effective at pumping water vapor into the high atmosphere, compared to tropical convection on a rapidly rotating world.  However, there is much more work left to be done.  Here, our model uses a relatively low model top, with poor resolution of the stratosphere.  Furthermore, Yang et al. (2014), using CAM3 with an identical vertical resolution, found much lower water vapor mixing ratios in the upper atmosphere for an early (wet) Venus around the Sun.  Future planned work will utilize a higher model top, with significantly improved vertical resolution, and will study a wider range of stellar fluxes.

\subsection{Habitability of M-dwarfs}

Several studies have argued that terrestrial planets within the HZs of M dwarfs may not be habitable, either because they could 
be deficient in volatiles \citep{Lissauer2007} or form dry as a result of inefficient water delivery \citep{Raymond2007}, 
if they formed {\it in situ}. If the planets migrate from the outer parts of the system, they 
accumulate massive H and He envelopes.  
 Whether they are able to shed most of their H-rich envelopes and potentially become 'habitable evaporated
cores' (HECs) depends on the mass of the solid core and the X-ray/UV-driven escape \citep{Lugeretal2015}. Recently, 
\cite{LB2015} pointed out that due to the extended pre-main sequence phase of M dwarfs ($\sim 100$Myr, \cite{Baraffe2015}),
where the star is more luminous than its main sequence phase, terrestrial planets in the (current) HZs of M dwarfs
could have been in runaway greenhouse during the pre-main sequence phase and may therefore be uninhabitable. This
 prolonged runaway greenhouse state may 
have removed water from a planet's atmosphere through photolysis, and may dessicate the planet. As a by-product, if the 
surface of the planet is not an efficient oxygen sink, the resulting oxygen from the photolysis may accumulate and 
can potentially generate an abiotically produced bio-signature \citep{LB2015}.
However, based on the combined results from planetary population synthesis and atmospheric water-loss models, 
\cite{TI2015} suggest a bimodel distribution of water contents for terrestrial planets around low mass stars. 
 Earth-like water endowments may be rare, but dune and deep ocean worlds may both be common.  Thus, the 
extended pre-main sequence phase of M-dwarfs does not  unequivocally rule out the existence of ocean planets.

\section{Conclusions}
\label{conclusions}
Low-eccentricity planets at the inner edge of the HZ around low-mass stars that are within the tidal locking
 radius and not perturbed by planetary companions 
are expected to be in synchronous rotation, with $1:1$ 
spin-orbit resonance. 
In this paper, we showed that to determine the inner edge of the HZ around low-mass stars using 3-D GCMs, care 
should be taken to self-consistently calculate the orbital/rotational periods of the planets 
as a function of the incident stellar flux, in accordance with
Kepler's third law. Previous GCM studies that assumed fixed, 60-day orbital periods were always in the slow-rotator 
dynamical regime, and thus may have overestimated fractional cloud cover and planetary albedo on planets near the inner 
edge of the HZ. According to our model, planets orbiting late-M dwarfs should begin 
transitioning to the fast-rotator regime, 
resulting in zonal banding and a decrease in cloud cover, with a corresponding decrease in planetary albedo. 
Thus, in our model the HZ 
inner edge moves outward for planets orbiting late M dwarfs compared to previous calculations. For hotter stars, though, ($T_{eff} > 3700$ K), the inner 
edge moves inwards as a result of improved numerics in the CAM4 GCM. 

We also showed that, for a given stellar $T_{eff}$, the location of the HZ for M-dwarfs 
depends on the metallicity of the star, because metallicity affects the 
stellar luminosity and 
hence, the semi-major axis and orbital period of a planet near the inner edge. Low-metallicity stars are more luminous
 for a given $T_{eff}$, and hence their HZ boundaries are farther out.

Future work will include updating the radiative transfer scheme in CAM4, and taking into account that water can become 
major atmospheric constituent  on warm, moist planets. Still, 
 these results are a first step towards improved estimates of the HZ inner edge. 
We encourage model intercomparisons, and all 3-D calculations of HZ boundaries should be validated by several independent  GCMs, as is done routinely in the global climate change community.

%\begin{figure}[!hbp|t]
%\includegraphics[width=.85\textwidth]{Lum_Teff.eps}
%\caption{Caption}
%\label{lumteff}
%\end{figure}

%	A FORTRAN code is available with the online version of the paper. An interactive webpage to obtain HZs is available at:
%        \url{http://www3.geosc.psu.edu/~ruk15/planets/} or at 
%	\url {http://depts.washington.edu/naivpl/content/hz-calculator}. 

	\acknowledgements

        The authors would like to thank  Daniel Koll and Dorian Abbot for kindly providing the aquaplanet patch
        for CESM, and responding to our inquiries,  that enabled us to accomplish this work. 
        The authors appreciate constructive comments and suggestions from an anonymous reviewer that improved the
        manuscript. The authors also thank Michael Way and Tony Del Genio from NASA GISS for providing detailed 
        comments on an earlier
        version of the manuscript.
        R. K, J.F.K and V. M gratefully acknowledge funding from NASA Astrobiology
        Institute's  Virtual 
       Planetary Laboratory lead team, supported by NASA under cooperative agreement
       NNH05ZDA001C.
       J.H.-M. acknowledges support from the Virtual Planetary Laboratory under award NNX11AC95G,S03.
       E.T.W thanks NASA Planetary Atmospheres Program award NNH13ZDA001N-PATM. 
       S. M and R. T acknowledge support from NSF grants AST 1006676, AST 1126413, and AST 1310885.
       This work was partially supported by funding from the Center for Exoplanets and Habitable Worlds. The Center for Exoplanets and Habitable Worlds is supported by the Pennsylvania State University, the Eberly College of Science, and the Pennsylvania Space Grant Consortium. This work was also partially supported by the Penn State Astrobiology Research Center and the National Aeronautics and Space Administration (NASA) Astrobiology Institute.
	The authors acknowledge the Research Computing and Cyberinfrastructure
	unit of Information Technology Services at The Pennsylvania State
	University for providing advanced computing resources and services that
	have contributed to the research results reported in this paper. {\url {
	http://rcc.its.psu.edu}}.  This work was also facilitated through the use of 
	advanced computational, storage, and networking infrastructure provided by the 
	Hyak supercomputer system, supported in part by the University of Washington eScience Institute.
        This work also utilized the Janus supercomputer, which is supported by the National Science Foundation 
        (award number CNS-0821794) and the University of Colorado at Boulder .
%	This research has made use of the Exoplanet Orbit Database
%	and the Exoplanet Data Explorer at exoplanets.org.

%        The authors thank an anonymous reviewer whose comments greatly improved the manuscript.
%	R. K, R. R, J.F.K and S.D.G gratefully acknowledge funding from NASA Astrobiology
%	 Institute's  Virtual 
%	Planetary Laboratory lead team, supported by NASA under cooperative agreement
%	NNH05ZDA001C, and the Penn State Astrobiology Research Center.
%	V.E. acknowledges the support of the European Research Council (Starting Grant 209622: E3ARTHs).
%The Center for Exoplanets and Habitable Worlds is supported by the
%	Pennsylvania State University, the Eberly College of Science, and the
%	Pennsylvania Space Grant Consortium. R.K and R.R contributed equally to this work.


\begin{thebibliography}{}


%        \bibitem[Allard et al.(2003)]{Allard2003}
%Allard, F., Guillot, T., & Ludwig, H. G. et al. 2003, Brown Dwarfs (IAU Symp. 211), ed. E. Martín (San Francisco, CA: ASP), 325

%\bibitem[Allard et al.(2007)]{Allard2007}
%Allard, F., Allard, N. F., \& Homeier, D. et al. 2007, \aap, 474, L21

%
        \bibitem[Anglada-Escude et al.(2013)]{Anglada-Escude2013}        
        Anglada-Escude, G., Tuomi, M., Gerlach, E. et al. 2013. \aap, 556, id.A126

	
        \bibitem[Baraffe et al.(2015)]{Baraffe2015}
        Baraffe, I., Homeier, D., Allard, F., \& Chabrier, G. 2015. \aap, 577, id.A42




   \bibitem[Boyajian et al.(2012)]{Boyajian2012}
    Boyajian, T. S., Von Braun, K., Van Belle, G., et al. 2012. \apj, 757, 112

   \bibitem[Boyajian et al.(2014)]{Boyajian2014}
    Boyajian, T. S., Van Belle, G., Braun, V. et al. 2014. \aj, 147, 47
   
\bibitem[Carone et al.(2014)]{Carone2014}
Carone L., Keppens, R., \& Decin, L. (2014). MNRAS, 445, 930

        \bibitem[Chabrier \& Baraffe(2000)]{ChabBara2000}
         Chabrier, G., \& Baraffe, I. 2000. ARAA, 38, 337



        \bibitem[Dotter et al.(2008)]{Dotter2008}
        Dotter, A., Chaboyer, B., Jevremovic, D., et al. 2008. \apjs, 178, 89

%        \bibitem[Dressing \& Charbonneau(2013)]{DC2013}
%         Dressing, C., \& Charbonneau, D. 2013. \apj, 767, 1
 

        \bibitem[Edson et al.(2011)]{Edson2011}
         Edson, A., Lee, S., Bannon, P. et al. 2011. Icarus, 212, 1


%\bibitem[Fortney et al.(2007)]{Fortney2007}
%Fortney, J. J., Marley, M. S., Barnes, J. W. 2007. \apj, 659, 1661

%        \bibitem[Gaidos(2013)]{Gaidos2013}
%         Gaidos, E. 2013, \apj, 770, 90

\bibitem[Haqq-Misra \& Kopparapu(2015)]{HK2015}
Haqq-Misra, J., \& Kopparapu, R. K. 2015. MNRAS. 446, 428

\bibitem[Heng \& Vogt(2011)]{HV2011}
Heng, K., Vogt, S. 2011, MNRAS, 415, 2145

\bibitem[Hunten(1973)]{Hunten1973}
Hunten, D. M. 1973. J. Atmos. Sci., 30, 1481

\bibitem[Joshi et al.(1997)]{Joshi1997}
Joshi, M. M., Haberle R. M., Reynolds R. T. 1997. {\it Icarus}, 129, 450

\bibitem[Kaspi \& Showman(2015)]{KS2015}
Kaspi, Y., Showman, A. P. 2015. \apj, 804, 60

	\bibitem[Kasting et al.(1993)]{Kasting1993}
	Kasting, J., F., Whitmire, D., P., \& Reynolds. R. T. 1993, {\it Icarus}, 101, 108


	\bibitem[Kopparapu et al.(2013)]{Kopp2013}
	Kopparapu, R. K., Ramirez, R., Kasting, J. F., Eymet, V., Robinson, T. D., Mahadevan, S.,
	Terrien, R. C., Domagal-Goldman, S. D., Meadows, V., \& Deshpande, R. 2013, \apj, 765, 131

%        \bibitem[Kopparapu(2013)]{KoppM2013}
%         Kopparapu, R. K. 2013. \apjl, 767, 1

   \bibitem[Kopparapu et al.(2014)]{Kopp2014}
    Kopparapu, R. K., Ramirez, R. M., SchottelKotte, J. et al. 2014. \apjl, 787, L29

\bibitem[Leconte et al.(2013)]{Leconte2013}
Leconte, J., Forget, F., Charnay, B., Wordsworth, R., \& Pottier, A. 2013. Nature, 504, 268

\bibitem[Leconte et al.(2015)]{Leconte2015}
Leconte, J., Wu, H., Menou, K., \& Murray, N. 2015. Science, 347, 632

\bibitem[Lissauer(2007)]{Lissauer2007}
Lissauer, J. J. 2007. \apjl, 660, L149

\bibitem[Luger \& Barnes(2015)]{LB2015}
Luger. R., \& Barnes, R. 2015. Astrobiology, 15, 2

\bibitem[Luger et al.(2015)]{Lugeretal2015}
Luger. R., Barnes, R., Lopez, E., Fortney, J., Jackson, B., \& Meadows, V. 2015. Astrobiology, 15, 57

%\bibitem[Lopez \& Fortney(2013)]{LF2013}
%Lopez, E. D., \& Fortney, J. J. 2013, \apj submitted, arXiv:1311.0329

\bibitem[Mann et al.(2015)]{Mann2015}
Mann, W. A., Feiden, A. G., Gaidos, E., Boyajian, T., \& Von Braun, K. 2015. \apj, 804, 64 

\bibitem[Neale et al.(2010)]{Neal2010}
Neale, R. B., Gettelman, S. P., Chen, C. C. et al. 2010. NCAR/TN-486+STR NCAR TECHNICAL NOTE

\bibitem[Newton et al.(2015)]{Newton2015}
Newton, E., Charbonneau, D., Irwin, J., \& Mann, A. 2015. \apj, 800, 85


%\bibitem[Pierrehumbert \& Gaidos(2011)]{PG2011}
%Pierrehumbert, R. T., \& Gaidos, E. 2011. \apjl, 734, L13

\bibitem[Ricker et al.(2014)]{Ricker2014}
Ricker, G. R., Winn, J. N., Vanderspek, R. et al. 2014. Proceedings of the SPIE, 9143, 914320


\bibitem[Raymond et al.(2007)]{Raymond2007}
Raymond, S. N., Scalo, J., Meadows, V. S. 2007. \apj, 669, 606

\bibitem[Robertson et al.(2014)]{Robertson2014}
Robertson, P., \& Mahadevan, S. 2014. \apjl, 793, L24


\bibitem[Selsis et al.(2007b)]{Selsis2007b}
Selsis, F. et al. 2007b. \aap, 476, 137

%\bibitem[Pierrehumbert(2010)]{RayP2010}
%Pierrehumbert, R. T. 2010. Principles of Planetary Climate, Cambridge University Press

\bibitem[Sherwood \& Huber(2010)]{SH2010}
Sherwood, S. C. \& Huber, M. 2010. PNAS, 107(21), 9552


\bibitem[Showman et al.(2013a)]{Showman2013a}
Showman, A. P.,Fortney. J. J., Lewis, N. K., Shabram, M. 2013. \apj, 762, 24  

\bibitem[Showman et al.(2013)]{Showman2013}
Showman, A.P., R.D. Wordsworth, T.M. Merlis, and Y. Kaspi. 2013. Comparative Climatology of Terrestrial Planets (S.J. Mackwell et al., Eds.), Univ. Arizona Press. pp. 277-326.

\bibitem[Terrien et al.(2015)]{Terrien2015}
Terrien, R. C., Mahadevan, S., Bender, C. F., Deshpande, R., \& Robertson, P. 2015. \apjl, 802, L10

\bibitem[Tian \& Ida(2015)]{TI2015}
Tian, F., \& Ida, S. Nature Geoscience, 8, 177

\bibitem[Way et al.(2015)]{Way2015}
Way, M. J., Del Genio, A. D., Kelley, M., Aleinov, I., Clune, T. 2015. Comparative Climatology of Terrestrial Planets II, NASA Conference Proceeding, arxiv:1511.07283

\bibitem[Wolf \& Toon(2014)]{WT2014}
Wolf, E., \& Toon, E. 2014. Geophysical Research Letters. 41, doi:10.1002/2013GL058376

\bibitem[Wolf \& Toon(2015)]{WT2015}
Wolf, E., \& Toon, E. 2015. JGRA, 120, 5775

\bibitem[Wordsworth \& Pierrehumbert(2013)]{WP2013}
Wordsworth, R., \& Pierrehumbert, R. 2013. Science, 339, 64

\bibitem[Yang et al.(2013)]{Yang2013}
Yang, J., Cowan, N. B., \& Abbot, D. S. 2013. \apjl, 771, L45

\bibitem[Yang et al.(2014a)]{Yang2014a}
Yang, J., Gwenael,  B., Fabrycky, D., \& Abbot, D.  2014. \apjl, 787, L2

\bibitem[Young et al.(2012)]{YoungP2012}
Young, P. A., Liebst, K., \& Pagano, M. 2012. \apjl, 755, L31

\end{thebibliography}
\end{document}